\documentclass{article}

\usepackage[utf8]{inputenc}
\usepackage[T1]{fontenc}

\usepackage{graphicx} % Required for inserting images
\usepackage{amsmath}
\usepackage{amssymb}

\newif\ifFull
\Fulltrue

\newif\ifBlind
\Blindfalse

\newcommand{\Full}[2][]{\ifFull#2\else#1\fi}
\newcommand{\Short}[2][]{\Full[#2]{#1}}

\newcommand{\Open}[2][]{\ifBlind#1\else#2\fi}

\ifFull
\def\shortTab{}
\else
\def\shortTab{&}
\fi

\usepackage{epsfig}
\usepackage{subcaption}
\usepackage{calc}
\usepackage{amstext}
\usepackage{amsthm}
\usepackage{multicol}
\usepackage{pslatex}
\usepackage[ruled,lined,noend]{algorithm2e}
\usepackage[bottom]{footmisc}
\usepackage{xcolor}

\ifFull

\usepackage[
  ,a4paper
]{geometry}

\usepackage{etoolbox}
\makeatletter
\patchcmd{\maketitle}{\@fnsymbol}{\@alph}{}{}
\makeatother

\usepackage[
    ,colorlinks
    ,linkcolor=blue
]{hyperref}

\newcommand{\Uppercase}[1]{#1}

\newcommand{\authorname}[1]{#1}
\newcommand\orcidAuthor[1]{
\hspace*{-0.5mm}%
\epsfig{file=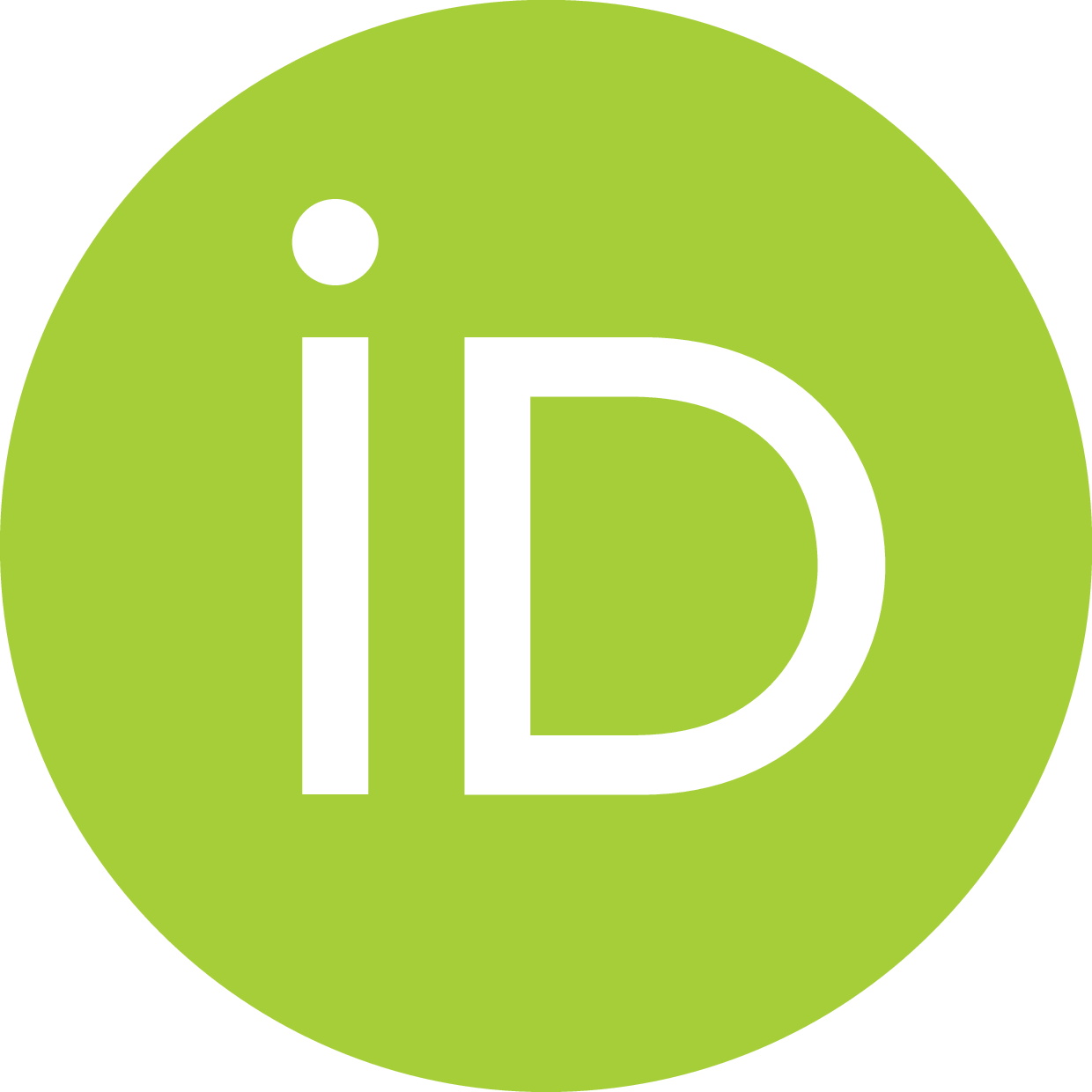,width=0.3cm}\footnote{\protect\epsfig{file=orcid.eps,width=0.3cm}~https://orcid.org/#1}%
\hspace*{0.5mm}
}
\newcommand{\affiliation}[1]{\\\emph{#1}}
\newcommand{\email}[1]{\\\emph{#1}}
\newcommand{\keywords}[1]{\def\mykeywords{#1}}
\renewcommand{\abstract}[1]{\def\myabstract{#1}}
\date{}

\renewcommand{\sup}[1]{\ensuremath{^{\text{#1}}}}

\else
\usepackage{apalike}
\usepackage[
    ,colorlinks
    ,citecolor=blue
    ,linkcolor=blue
]{hyperref}
\usepackage{SCITEPRESS}     % Please add other packages that you may need BEFORE the SCITEPRESS.sty package.

\newcommand{\Uppercase}[1]{\uppercase{#1}}
\fi %% ifFull

\newtheorem{definition}{Definition}
\newtheorem{theorem}{Theorem}
\newcommand{\MAPFR}{$\mathrm{MAPF}_R$}
\newcommand{\MAPFRSym}{P}
\newcommand{\PlanSym}{\Pi}
\newcommand{\Cost}[1]{\mathit{cost}(#1)}
\newcommand{\isCollision}{\ensuremath{\textsc{IsCollision}}}
\newcommand{\collGen}{\varphi_{c_{\mathit{new}}}}
\newcommand{\LRA}{\ensuremath{\mathsf{QF\_LRA}}}

\newcommand{\smtCcbs}{SMT-CCBS}
\newcommand{\ccbs}{CCBS}
\newcommand{\smtLra}{SMT-LRA}

\newcommand{\New}[1]{\textcolor{red}{#1}}

\newcommand{\at}[2][j]{\ensuremath{ #2^{[#1]} }}

\title{Multi-Agent Path Finding with Continuous Time Using\\SAT Modulo Linear Real Arithmetic}

\author{\Open{\authorname{Tomáš Kolárik\sup{1,2}\orcidAuthor{0000-0002-7207-5197}, Stefan Ratschan\sup{1}\orcidAuthor{0000-0003-1710-1513} and Pavel Surynek\sup{2}\orcidAuthor{0000-0001-7200-0542}}
\Full{\smallskip}
\affiliation{\sup{1}Institute of Computer Science of the Czech Academy of Sciences}
\affiliation{\sup{2}Faculty of Information Technology, Czech Technical University in Prague}
\Full{\smallskip}
\email{\{kolarik,stefan.ratschan\}@cs.cas.cz}
\Short{\vspace*{-3ex}}
\email{\{tomas.kolarik,pavel.surynek\}@fit.cvut.cz}
}}

\keywords{Multi-Agent Path Finding, Satisfiability Modulo Theories}

\abstract{This paper introduces a new approach to solving a continuous-time version of the multi-agent path finding problem. The algorithm translates the problem into an extension of the classical Boolean satisfiability problem, satisfiability modulo theories (SMT), that can be solved by off-the-shelf solvers. This enables the exploitation of conflict generalization techniques that such solvers can handle. Computational experiments show that the new approach scales better with respect to the available computation time than state-of-the art approaches and is usually able to avoid their exponential behavior on a class of benchmark problems modeling a typical bottleneck situation.}

\begin{document}

\ifFull
\maketitle
\medskip

\textbf{Keywords: }\mykeywords
\bigskip
\medskip

{\centering \bf \large Abstract\\}
\smallskip
\myabstract
\else
\onecolumn \maketitle \normalsize \setcounter{footnote}{0} \vfill
\fi

\section{\Uppercase{Introduction}}

Multi-agent path finding (MAPF) \cite{Silver05,Ryan08,LunaB11,WildeMW14} is the problem of navigating agents from their start positions to given individual goal positions in a shared environment so that agents do not collide with each other. The standard discrete variant of the MAPF problem is modeled using an undirected graph in which $k$ agents move instantaneously between its vertices. The space occupancy by agents is modeled by the requirement that at most one agent reside per vertex and via movement rules that forbid conflicting moves that traverse the same edge in opposite directions.

Standard discrete MAPF however lacks expressiveness for various real life problems where continuous time and space play an important role such as robotics applications and/or traffic optimization \cite{FelnerSSBGSSWS17,Ma22}.

%\footnote{Pavel, I could not find any direct reason for the necessity of continuous time in those articles. In my opinion, we need a  100\% fool-proof explanation of the reason why continuous-time is important (i.e., some killer application), in order for this work to have impact.}

This drawback of standard MAPF has been mitigated by introducing various generalizations such as MAPF with continuous time (\MAPFR{}) \cite{AndreychukYSAS22}. This allows more accurate modeling of the target application problem without introducing denser and larger discretizations. Especially in applications, where agents correspond to robots, it is important to consider graph edges that interconnect vertices corresponding to more distant positions. It is unrealistic to consider unit time for such edges as done in the standard MAPF, hence general duration of actions must be adopted. The action duration often corresponds to the length of edges which implies fully continuous reasoning over the time domain.

In this paper, we show how to solve the \MAPFR{} problem by directly translating it to a satisfiability modulo theories (SMT)~\cite{Barrett:18} problem. SMT extends the classical Boolean satisfiability problem to formulas whose atomic sub-formulas may not only be formed by propositional variables, but also by first-order predicate language formulas whose syntax is restricted to certain predicate and function symbols and whose semantics by interpreting those predicate and function symbols according to a given first-order theory. This allows the exploitation of decision procedures for such theories by so-called SMT solvers. Typical theories include the linear theory of integers, the theory of bit-vectors, and the theory of free function symbols. In this paper, we will use the theory of quantifier free linear real arithmetic (\LRA{}), which will allow us to reason about time in MAPF modeled in a continuous manner.

State-of-the-art approaches for \MAPFR{} such as Continuous-time Conflict-based Search (\ccbs{})~\cite{AndreychukYSAS22}, a generalization of Conflict-based Search (CBS)~\cite{SharonSFS15} that represents one of the most popular algorithms for MAPF, search for optimal plans. However, in real-world applications, where the formalized MAPF problem results from an approximation of the original application problem, an overly strong emphasis on optimality is often pointless. Moreover, it may result in non-robust plans that are difficult to realize in practice \cite{AtzmonSFSK20}. Hence we aim for a sub-optimal method whose level of optimality can be adapted to the needs in the given application domain.

Unlike methods based on \ccbs{} that approaches the optimum from below by iterating through plans that still contain collisions, our method approaches the optimum from above, iterating through collision-free plans. This has the advantage that---after finding its first plan---our method can be interrupted at any time, still producing a collision-free, and hence feasible plan. This anytime behavior is highly desirable in practice \cite{LiCHSK21}.

Another advantage over existing methods is the fact that the objective function is a simple expression handed over to the underlying SMT solver. This allows any objective function than the SMT solver is able to handle without the need for any algorithmic changes.

%    \item Various objectives can be integrated easily. To change the objective from makespan to sum-of-costs it is only needed change %the definition of the variable being optimized.

We did experiments comparing our method with the state-of-the art approaches \ccbs{} and \smtCcbs~\cite{AndreychukYSAS22} on three classes of benchmark problems and various numbers of agents. The results show that our method is more sensitive to time-outs than the existing approaches, typically being able to solve more instances than existing approaches for high time-outs and less for lower time-outs. Future improvement of computer efficiency will consequently make the method even more competitive.

Moreover, for one class of benchmark problems---modeling a bottleneck situation where all agents have to queue for passing a single node, the new method usually avoids the exponential behavior of \ccbs{} and \smtCcbs{} whose run-times explode from a certain number of agents on. Such bottleneck situations frequently occur for certain types of application problems (e.g., in traffic problems or navigation of characters in computer games through tunnels and the like).

\paragraph{Further Related Work: }
Existing methods for generalized variants of MAPF with continuous time include variants of Increasing Cost Tree Search (ICTS) \cite{WalkerSF18} where durations of individual actions can be non-unit. The difference from our generalization is that agents do not have an opportunity to wait an arbitrary amount of time but wait times are predefined via discretization. Similar discretization has been introduced in the Conflict-based Search algorithm \cite{0002UKK19}. Since discretization in case of ICTS as well as CBS brings inaccuracies of representation of the time, it is hard to define optimality. Moreover, a more accurate discretization often increases the number of actions, which can lead to an excessively large search space.

Our method for \MAPFR{} comes from the stream of compilation-based methods for MAPF, where the MAPF instance is compiled to an instance in a~different formalism for which an off-the-shelf efficient solver exists. Solvers based on formalisms such as Boolean satisfiability (SAT) \cite{SurynekFSB16,Surynek19}, Answer Set Programming \cite{BogatarkanP019}, Constraint Programming (CP) \cite{Ryan10,GangeHS19}, or Mixed-integer Linear Programming (MILP) \cite{LamBHS22} exist. The advantage of these solvers is that any progress in the solver for the target formalism can be immediately reflected in the MAPF solver that it is based on.

The earlier MAPF method related to SMT (the SMT-CBS algorithm)~\cite{Surynek19} separates the rules of MAPF into two logic theories, one theory for conflicts between agents and one theory for the rest of the MAPF rules. The two theories are used to resolve conflicts between agents lazily similarly as it is done by the CBS algorithm.

%\footnote{Pavle, mohl bys prosím ten rozdíl k přístupu v současném článku ještě popsat trochu podrobněji?}

The application of SAT and SMT solvers to planning problem different from MAPF is not new~\cite{Rintanen:21,Leofante:23,Cashmore:20}, usually in the context of temporal and numerical planning---extensions of the classical planning problem with numerical variables. SMT solvers have been used for specific planning problems with multiple agents~\cite{Kolarik:23}, employing however a synchronous model that  identifies each step of the unrolled planning problem with a fixed time period.
% Railway route planning can also be viewed
% as a~multi-agent path finding problem~(Train Route Planning as a Multi-agent Path Finding Problem, 2021),
% where trains are viewed as agents.

\paragraph{Acknowledgments.}
The work of Tomáš Kolárik was supported by the project 22-31346S of the Czech Science Foundation GA ČR  and by CTU project SGS20/211/OHK3/3T/18.
The work of Stefan Ratschan was supported by the project 21-09458S of the Czech Science Foundation GA ČR  and institutional support RVO:67985807.
The work of Pavel Surynek was supported by the project 22-31346S of the Czech Science Foundation GA ČR.

\section{\Uppercase{\MAPFR{}: Problem Definition}}
\label{sec:problem}

We follow the definition of multi-agent path finding with continuous time (\MAPFR{}) from \cite{AndreychukYSAS22}.

We define a \MAPFR{} problem by the tuple $(G,M,A,s,g,\mathit{coord})$, where $G=(V,E)$ is a~directed graph with $V$ modeling important positions in the environment and $E$ modeling possible transitions between the positions, $M$ is a metric space that models the continuous environment, $A=\{a_1,a_2,...,a_k\}$ is a~set of agents, functions $s:A \rightarrow V$ and $g:A \rightarrow V$ define start and goal vertices for the agents, and $\mathit{coord}:V \rightarrow M$ assigns each vertex a coordinate in metric space $M$.

The edges $E$ define a set of possible move actions, where each $e = (u,v) \in E$ is assigned a duration $e_D\in\mathbb{R}_{>0}$ and a motion function $e_M: [0,e_D] \rightarrow M$ where $e_M(0) = \mathit{coord}(u)$ and $e_M(e_D) = \mathit{coord}(v)$.
%% in the future, it migth be necessary to use multigraph (e.g. multiple geometric curves)
In addition to this, there is infinite set of wait actions associated with each vertex $v \in V$ such that an agent can wait in $v$ any amount of time. The motion function for a wait action is constant and equals to $\mathit{coord}(v)$ throughout the duration of the action.

Collisions between agents are defined via a~collision-detection predicate $\isCollision{}\subseteq A \times A \times M \times M$ such that $\isCollision{}(a_i,a_j,m_i,m_j)$ if and only if the bodies of agents $a_i$ and $a_j$ overlap at coordinates $m_i$ and $m_j$. For this purpose, we assume that the bodies are open sets and overlapping is understood to be strict. Hence agents are permitted to touch if they are assumed to have a closed boundary which is not defined as a collision.

The algorithm described in this paper is abstract in the sense that it does not explicitly restrict the class of motion actions.
Instead it assumes that it is possible to do collision detection and avoidance, as described in Section~\ref{sec:coll}. This
is possible, for example, if the agents and motion functions are described by polynomials, due to the fact that the theory of real closed fields allows quantifier elimination. Note that this allows the modeling of non-constant agent speed and of movements along non-linear curves. Still, in our implementation, for reasons of efficiency and ease of implementation, the motion functions are required to be linear.

%\footnote{ This gives an impression that all actions are related to edges, but wait actions are not. I would prefer for example $\pi=(\alpha_1,\alpha_2,...,\alpha_n)$ }

% Zkusil jsem zmenu na alpha a nedelal bych to. Cekaci akce muze byt chapana jako odvozena od smycky ve vrcholu. Bylo by to zbytecne zavadeni symbolu, kterych je tam az az

Given a sequence of actions $\pi=(e_1,e_2,...,e_n)$, we generalize the duration and motion functions from individual actions to overall $\pi$ which we denote by $\pi_D$ and $\pi_M$, respectively. Let $\pi[:{n'}]=(e_1,e_2,...,e_{n'})$ denote the prefix of the sequence of actions, then $\pi_D=\sum_{i=1}^{n}{{e_i}_D}$ and analogously $\pi[:{n'}]_D=\sum_{i=1}^{n'}{{e_i}_D}$. The motion function $\pi_M$ needs to take into account the relative time of individual motion functions ${e_i}_M$, that is: $\pi_M(t)={e_1}_M(t)$ for $t \leq {e_1}_D$, ..., $\pi_M(t)={e_{n'}}_M(t-\pi[:{n'-1}]_D)$ for $\pi[:{n'-1}]_D \leq t \leq \pi[:{n'}]_D$, ..., $\pi_M(t)={e_{n}}_M({e_{n}}_D)$ for $\pi_D < t$. The last case means that the agent stops after executing the sequence of actions and stays at the coordinate of the goal vertex.

\begin{definition}
There is a collision between sequences of actions $\pi_i$ and $\pi_j$ if and only if $\exists t \in [0,\allowbreak \max\{{\pi_i}_D,\allowbreak{\pi_j}_D\}]$ such that $\isCollision{}(a_i,a_j,{\pi_i}_M(t),{\pi_j}_M(t))$.
\end{definition}

\begin{definition}
A \emph{pre-plan} of a given \MAPFR{} problem $(G,M,A,s,g,\mathit{coord})$ is a collection of sequences of actions $\pi_1$, $\pi_2$, ..., $\pi_k$ s.t. for every $i\in\{ 1,\dots, k\}, \pi_i(0)=s(a_i)$ and $\pi_i({\pi_i}_D)=g(a_i)$.
A \emph{plan} for given \MAPFR{} problem $\MAPFRSym$ is a pre-plan of $\MAPFRSym$ whose sequences are pair-wise collision free.
\end{definition}

We define several types of cost functions that we denote by $\Cost{\PlanSym}$, for a given plan $\PlanSym$. For example, we will work with
sum-of-costs (in this case, $\Cost{\PlanSym}=\sum_{i=1}^{k}{\Cost{{\pi_i}_D}}$), or makespan.

For a \MAPFR{} problem $\MAPFRSym$, we denote by $opt(\MAPFRSym)$ its optimal plan and by $opt_{pre}(\MAPFRSym)$ its optimal pre-plan. Clearly  $\Cost{opt_{pre}(\MAPFRSym)}\leq \Cost{opt(\MAPFRSym)}$, but $opt_{pre}(\MAPFRSym)$ is much easier to compute than $opt(\MAPFRSym)$ since it  directly follows from the plans of the individual agents.

For our approach, the following two observations are essential:
\begin{itemize}
    \item Multiple subsequent wait actions can always be merged into a single one without changing the overall motion.
    \item It is always possible to insert a wait action of zero length between two subsequent move actions---again without changing the overall motion.
\end{itemize}
Due to this, we can restrict the search space to plans for which each wait action is immediately followed by a move action and vice versa. Our SMT encoding will then be able to encode wait and move actions in pairs, which motivates counting the number of steps of plans by just counting move actions. Hence, for a~sequence of actions $\pi$ we denote by $|\pi|$ the number of move actions in the sequence, and for a plan $\PlanSym$, we call
 $|\PlanSym|:=max_{i=1}^{k}{|\pi_i|}$ the number of steps  of plan $\PlanSym$.

\section{\Uppercase{Algorithm}}
\label{sec:alg}

Our goal is to encode the planning problem in an SMT theory that is rich enough to model time and to represent conflict generalization constraints. Since SMT solvers only encode a fixed number of steps, we have to use a notion of optimality that takes this into account. Hence the first optimization criterion is the number of steps, and the second criterion cost, which we optimize up to a given $\delta>0$:

\begin{definition}
  \label{def:opt}
  %Given a \MAPFR{} problem $\MAPFRSym$, $\mathit{minsteps}(\MAPFRSym)$ is the minimal $h\geq |opt_{pre}(P)|$ such that $\MAPFRSym$ has a
 %a plan $\Pi$ with $|\Pi|=h$.

 A plan $\PlanSym$ satisfying a \MAPFR{} problem $\MAPFRSym$ is
  \emph{minstep $\delta$-optimal} iff
  \begin{itemize}
  \item $|\PlanSym|=\min\{ |\Pi| \mid \Pi\text{ is a plan of } P \}$, and
  \item $\Cost{\PlanSym}\leq (1+\delta)\inf\{\Cost{\Pi'}\mid |\Pi'|=|\Pi|\}$.
  %$\Cost{\PlanSym}\leq (1+\delta)\min_{|\Pi|=\mathit{minsteps}(\MAPFRSym)} \Cost{\Pi}$.
   %\mathit{optcost}_\Pi(\MAPFRSym)$ where $\mathit{optcost}_k(\MAPFRSym)$ is the minimal cost of plans of $P$ that take not more than $k$ steps and $\infty$ if such a plan does not exist.
  \end{itemize}
\end{definition}

The result is Algorithm~\ref{alg:main}. It searches from below for a plan of minimal number of steps, and then minimizes cost for the given number of steps using iterative bisection. %\footnote{Here it would also be possible to use optimizing SMT solvers, but $\dots$}.
For this, it uses a function $\mathit{findplan}$
that searches for a~plan with a~fixed number of states
whose cost is between some minimal and maximal cost
and that we will present in more details below in Algorithm~\ref{alg:findplan}.

\setlength{\algomargin}{0cm}

\SetKw{Let}{let}

\begin{algorithm}[!htb]
\caption{Main algorithm MAPF-LRA.}
\label{alg:main}
\DontPrintSemicolon
\BlankLine
  \begin{tabbing}\hspace*{0.3cm}\=\kill
    MAPF-LRA$(\MAPFRSym, \delta) \rightarrow p_{\mathit{opt}}$\\
    \KwIn{}\\
    \> - a \MAPFR{} problem $\MAPFRSym=(G,M,A,s,g,\mathit{coord})$\\
    \> - $\delta\in\mathbb{R}_{>0}$\\
    \KwOut{}\\
    \> - $p_{\mathit{opt}}$: a minstep $\delta$-optimal plan for $\MAPFRSym$\\
  \end{tabbing}
  \Full{\vspace*{-1ex}}
  \vspace*{-2ex}

  $h\leftarrow |opt_{pre}(P)|$\;
  $t_{\mathit{min}}\leftarrow \Cost{opt_{pre}(P)}$\;
  $C\leftarrow\emptyset$\;
  $(p,C)\leftarrow \mathit{findplan}(P,h,t_{\mathit{min}}, \infty, C)$\;
  \While{$p=\bot$}{
    $h\leftarrow h+1$\;
    $(p,C)\leftarrow \mathit{findplan}(P,h,t_{\mathit{min}}, \infty, C)$ \;
  }
  $p_{\mathit{opt}}\leftarrow$ $p$\;
  \While{$\Cost{p_{\mathit{opt}}}>(1+\delta) t_{\mathit{min}}$}{
    \Let $\hat{t}\in (t_{\mathit{min}}, \Cost{p_{\mathit{opt}}})$\;
    $(p,C)\leftarrow \mathit{findplan}(P,h,t_{\mathit{min}}, \hat{t}, C)$\;
    \leIf{$p=\bot$}{$t_{\mathit{min}}\leftarrow \hat{t}$}{$p_{\mathit{opt}}\leftarrow p$}
  }
  \Return{$p_{\mathit{opt}}$}
\end{algorithm}

When using a SAT solver to implement the function $\mathit{findplan}$, it would be an overkill to encode the \emph{whole} planning problem at once, since we would have to encode the avoidance of a huge number of potential collisions. Instead, we will encode this information on demand, initially looking for a pre-plan, and adding information on collision avoidance only based on collisions that have already occurred.

However, whenever a collision occurs, we do not only avoid the given collision, but also collisions that are in some sense similar. We will  call this a \emph{generalization} of a collision which we will also formalize in Section~\ref{sec:coll}.

So denote by $\varphi_{P,h,\underline{t},\overline{t}}$ an SMT formula encoding the existence of a  pre-plan $\Pi$ of \MAPFR{}-problem $\MAPFRSym$ with number of steps $h$ and cost in $[\underline{t},\overline{t}]$, that is, $|\Pi|=h$ and $\Cost{\Pi}\in [\underline{t},\overline{t}]$ (see Section~\ref{sec:encoding} for details).  We will use an SMT solver to solve those formulas and assume that for any formula $\varphi$ encoding a planning problem, $sat(\varphi)$ either returns the pre-plan satisfying $\varphi$ or $\bot$ if $\varphi$ is not satisfiable.

The result is Algorithm~\ref{alg:findplan} below.
\begin{algorithm}[!htb]
\caption{Function $\mathit{findplan}$ that searches for a~plan with a~bounded cost.}
\label{alg:findplan}
\DontPrintSemicolon
\BlankLine
  \begin{tabbing}\hspace*{0.5cm}\=\hspace*{0.5cm}\=\hspace*{0.5cm}\=\kill
    $\mathit{findplan}(h,t_{\min}, t_{\max}, C) \rightarrow (p, C')$\\
    \KwIn{}\\
    \> - $h\in\mathbb{N}_0$\\
    \> - $t_{\min}\in\mathbb{R}_{\geq 0}$\\
    \> - $t_{\max}\in\mathbb{R}_{\geq 0}\cup\{\infty\}$\\
    \> - $C$: a set of formulas that every plan must satisfy\\
    \KwOut{}\\
    \> - $p$: either a plan $\Pi$ with \Short{\\
                      \`}$|\Pi|=h$ and $\Cost{\pi}\in[t_{\min},t_{\max}]$, or\Short{\\
    \hspace*{1.0cm}} $\bot$, if such a plan does not exist\\
    \> - $C'$: a set of formulas that every plan must satisfy\\
  \end{tabbing}
  \Full{\vspace*{-1ex}}
  \vspace*{-3ex}

  $p\leftarrow sat(\varphi_{P,h,t_{\min},t_{\max}}\wedge\bigwedge_{\varphi_c\in C} \varphi_c)$\;
  \While{$\neg [p= \bot \vee p \text{ is collision-free}]$}{
    \Let $\collGen$ represent the generalization of \Short{\\\quad }collisions in $p$\;
    $C\leftarrow C\cup \{ \neg \collGen\}$\;
    $p\leftarrow sat(\varphi_{P,h,t_{\min},t_{\max}}\wedge\bigwedge_{\varphi_c\in C} \varphi_c)$\;
  }
  \Return{$(p, C)$}
\end{algorithm}

Note that if $p\neq\bot$, the pre-plan $p$ may have several collisions. The algorithm leaves it open for which of those collisions to add collision avoidance information into the formula $\collGen$. The algorithm leaves it open, as well, how much to generalize a found collision occurring at a certain point in time. In our approach, we use a specific choice here that we will describe in Section~\ref{sec:coll}.

   \begin{theorem}
     The main algorithm is correct, and if $\hat{t}$ is chosen as $(1-c)t_{\min}+c t_{\max}$, for some fixed $c\in (0,1)$, then it also terminates.
\end{theorem}

\begin{proof}
  Since $|opt_{pre}(\MAPFRSym)|$ is a lower bound on the number of steps of any plan of $P$, $h\leq\min\{ |\Pi| \mid \Pi\text{ is a plan of } P \}$ at the beginning of the first while loop.
  After termination of the first while loop, $h=\min\{ |\Pi| \mid \Pi\text{ is a plan of } P \}$. Moreover, the second while loop does not change $h$, and hence the result of the algorithm certainly satisfies the first condition of Definition~\ref{def:opt}.
  Throughout
  the first loop, $t_{\min}$ is a lower bound on all collision free plans,
  and throughout
  the second loop, it is a lower bound on all collision free plans that take $h$ steps, and $p_{opt}$ contains a $h$-step plan. Hence, after termination of the second loop also the second condition of Definition~\ref{def:opt} holds.

   Finally, if $c\in (0,1)$, $\Cost{p_{\mathit{opt}}}-t_{\min}$ goes to zero as the second-while loop iterates. Hence the termination condition of this loop must eventually be satisfied.
\end{proof}

\section{\Uppercase{SMT Encoding}}
\label{sec:encoding}

\newcommand{\defAg}{a}
\newcommand{\defAgB}{b}
\newcommand{\ag}[1][\defAg]{\ensuremath{#1}}
\newcommand{\ofAg}[2][\defAg]{\ensuremath{{#2}_{\ag[#1]}}}
\newcommand{\agB}{\ag[\defAgB]}
\newcommand{\ofAgB}[1]{\ofAg[\defAgB]{#1}}

\newcommand{\agVvar}[1][\defAg]{\ofAg[#1]{V}}
\newcommand{\agTvar}[1][\defAg]{\ofAg[#1]{T}}
\newcommand{\agWvar}[1][\defAg]{\ofAg[#1]{w}}
\newcommand{\agMvar}[1][\defAg]{\ofAg[#1]{m}}
\newcommand{\agVvarB}{\agVvar[\defAgB]}
\newcommand{\agTvarB}{\agTvar[\defAgB]}
\newcommand{\agWvarB}{\agWvar[\defAgB]}
\newcommand{\agMvarB}{\agMvar[\defAgB]}

\newcommand{\agTval}[1][\defAg]{\ofAg[#1]{\hat{T}}}
\newcommand{\agWval}[1][\defAg]{\ofAg[#1]{\hat{w}}}
\newcommand{\agMval}[1][\defAg]{\ofAg[#1]{\hat{m}}}
\newcommand{\agVvalB}{\agVval[\defAgB]}
\newcommand{\agTvalB}{\agTval[\defAgB]}
\newcommand{\agWvalB}{\agWval[\defAgB]}
\newcommand{\agMvalB}{\agMval[\defAgB]}

\newcommand{\costVar}{\ensuremath{\lambda}}

In this section, we present an~encoding
of the planning problem from Section~\ref{sec:problem} as an~SMT formula
in the quantifier-free theory of linear real arithmetic \LRA{}.
%\footnote{%
%It is not possible to use an~even simpler fragment of real arithmetic,
%difference logic (\RDL{}),
%because we require constraints that involve more than two variables.
%}.
Here we concentrate on the formula~$\varphi_{P,h,\underline{t},\overline{t}}$ that models time constraints of the agents
and their paths in graph~$G$ but does \emph{not} model collisions of the agents
and metric space~$M$. We leave the SMT encoding of collision avoidance to the next section.

\paragraph{Variables.}

As usual in planning applications of SAT solvers~\cite{Rintanen:21}, we unroll the planning problem in a~similar way as in Bounded Model Checking~\cite{Biere:09}, where each step $0, 1, \dots, h$ corresponds to one wait and one move action. As a~consequence, unrolling over $h$ steps corresponds to search for a pre-plan $\Pi$ with $|\Pi|=h$.

%A~variable~$x$ specific to a~discrete step~\(j \leq h\)
%has the form~\at{x}.
%The formula~$\varphi_{P,h,\underline{t},\overline{t}}$
%is built in an~incremental way---%
%new variables are introduced as late as~$h$ is increased.

Note that~$h$ corresponds to the maximum of the steps of all agents,
so an~agent that already reached the goal in step $j<h$
may remain in the same state in steps $j, \dots, h$,
without any further actions.

Each agent is modeled using a~separate set of Boolean and real-valued variables.
For each agent \(\ag{} \in A\) and discrete step~$j$,
we define variables~\at{\agVvar},
\at{\agTvar},~\at{\agWvar} and~\at{\agMvar}:
We model the vertex position of the agent by~\at{\agVvar}
which is a~Boolean encoding of a~vertex \(v \in V\)
using $\mathcal{O}(|V|)$ or $\mathcal{O}(\log(|V|))$ Boolean variables.
We will use the notation
\(\at{\agVvar} = v\)
to denote a~constraint that expresses that an~agent occupies vertex \(v \in V\)
at the beginning of discrete step~$j$.
We will also use \(\at{\agVvar} \neq v\)
as an~abbreviation for \(\neg\left(\at{\agVvar} = v\right)\).

Next, we model time constraints of the agent using real variables \at{\agTvar}, \at{\agWvar}, and \at{\agMvar}. The variables~\at{\agTvar} model
the absolute time when the agent occupies a~vertex that corresponds to~\at{\agVvar}, before it takes further actions within discrete step~$j$ (or later). The variables \at{\agWvar} model the duration of wait actions and the variables \at{\agMvar} the duration of move actions.

Finally, we use an auxiliary real variable~\costVar{} that, for a pre-plan~$\PlanSym$, corresponds to $\Cost{\PlanSym}$.
The objective is to minimize this variable.
There may be arbitrary linear constraints on the variable,
allowing specification of rich cost functions.

\paragraph{Constraints.}

We define (1) initial and goal conditions on the agents,
(2) constraints that ensure that the agents follow paths through the graph~$G$, and (3)
time constraints that correspond to occurrences of the agents at vertices of the graph.
For that, we only use the variables defined above.

The initial and goal conditions ensure that each agent $a$ visits the start and goal vertex
at the beginning and end of the plan, respectively:
\(\at[0]{\agVvar} = s(\ag) \land \at[h]{\agVvar} = g(\ag)\).

To ensure that the agents follow paths through the graph~$G$, we use for each agent $a$ and $j<h$
a~constraint ensuring that the pair of vertex positions~\at{\agVvar} and~\at[j+1]{\agVvar}
corresponds to an edge of the graph~$G$.
However, this is not necessarily the case
for an~already finished agent,
that is,
if \(\at{\agVvar} = g(\ag)\),
then also \(\at[j+1]{\agVvar} = g(\ag)\) is allowed.

The time constraints ensure that the initial value of time of all agents is zero:
\(\at[0]{\agTvar} = 0\).
For \(j > 0\),
they assume that during each discrete step, an~agent may first wait and then it moves,
resulting in the constraint
\(\at{\agTvar} = \at[j-1]{\agTvar} + \at[j-1]{\agWvar} + \at[j-1]{\agMvar}\).
For the waiting times, we require \(\at{\agWvar} \geq 0\).
In addition, we ensure that at least one agent starts to move
at the beginning of a~plan without waiting,
asserting $\bigvee_{\ag{} \in A} \at[0]{\agWvar} = 0$.
For the moving times, if $j<h$ we ensure that~\at{\agMvar}
corresponds to the duration of the edge between~\at{\agVvar} and~\at[j+1]{\agVvar}.
In addition,
\(\at{\agMvar} = 0\)
if \(j = h\)
or \(\at{\agVvar} = g(\ag) \land \at[j+1]{\agVvar} = g(\ag)\).

Note that agents are modeled asynchronously,
meaning that for a~pair of agents \(\ag{}, \agB{} \in A\),
\at{\agTvar} and~\at{\agTvarB{}}
corresponds \emph{not} necessarily to the same moment in time.
This implies that comparing times and the corresponding positions of agents,
in order to check whether there are collisions,
cannot be done in a~straightforward way,
and in the worst case,
variables corresponding to all discrete steps
must be examined.

We present two variants of cost functions:
sum of costs, defined as
\(\costVar{} = \sum_{\ag{} \in A} \at[h]{\agTvar}\),
and makespan, defined as
\(\costVar{} = \max_{\ag{} \in A} \at[h]{\agTvar}\).
To ensure that the formula~$\varphi_{P,h,\underline{t},\overline{t}}$
satisfies the bounds of the cost function,
we simply require
\(\costVar{} \geq \underline{t} \land \costVar{} \leq \overline{t}\).
An~example of an~alternative cost function is
\(\costVar{} = \sum_{\ag{} \in A} \sum_{j=0}^{h-1} \left( 2\at{\agMvar} + \at{\agWvar} \right)\)
which prefers minimizing moving times over waiting times
and can therefore result in more power-optimal plans.

Building the formula~$\varphi_{P,h,\underline{t},\overline{t}}$ from scratch after each increase of the number of steps $h$ would be inefficient. Hence we build the formula incrementally. However, some parts of the formula (e.g., the cost functions
or constraints such as \(\at[h]{\agVvar} = g(\ag)\)), explicitly depend on $h$, and hence need to be updated when $h$ is increased. Here we use the feature of modern SMT solvers, that allow the user to cancel constraints asserted after a previously specified milestone,
and to reuse the rest.

\section{\Uppercase{Collision Detection and Avoidance}}
\label{sec:coll}

\newcommand{\agR}[1][\defAg]{\ofAg[#1]{r}}
\newcommand{\agRB}{\agR[\defAgB]}

\newcommand{\Mof}[1]{#1_M}
\newcommand{\Dof}[1]{#1_D}
\newcommand{\aM}[1][\defAg]{\ofAg[#1]{\Mof{\alpha}}}
\newcommand{\aD}[1][\defAg]{\ofAg[#1]{\Dof{\alpha}}}
\newcommand{\aMB}{\aM[\defAgB]}
\newcommand{\aDB}{\aD[\defAgB]}

\newcommand{\agTcollVal}[1][\defAg]{\ofAg[#1]{\hat{\tau}}}
\newcommand{\agTcollValB}{\agTcollVal[\defAgB]}
\newcommand{\agTcollVar}[1][\defAg]{\ofAg[#1]{\tau}}
\newcommand{\agTcollVarB}{\agTcollVar[\defAgB]}

\newcommand{\isCollisionFull}[4]{\ensuremath{
  \isCollision{}(\ag[#1],\allowbreak \ag[#2],\allowbreak #3,\allowbreak #4)
}}
\newcommand{\isCollisionFullDef}[2]{\isCollisionFull{\defAg}{\defAgB}{#1}{#2}}

\newcommand{\inConflict}[2]{\ensuremath{\textsc{InConflict}_{\ag[#1], \ag[#2]}}}
\newcommand{\inConflictFull}[4]{\ensuremath{
  \inConflict{#1}{#2}(\aM[#1],\allowbreak \aM[#2],\allowbreak #3,\allowbreak #4)
}}
\newcommand{\inConflictDef}{\inConflict{\defAg}{\defAgB}}
\newcommand{\inConflictDefFull}[2]{\inConflictFull{\defAg}{\defAgB}{#1}{#2}}
\newcommand{\inConflictFullDef}{\inConflictDefFull{\agTcollVal{}}{\agTcollValB{}}}

\newcommand{\Safe}[2]{\ensuremath{\textsc{Safe}_{\ag[#1], \ag[#2]}}}
\newcommand{\SafeFull}[4]{\ensuremath{
  \Safe{#1}{#2}(\aM[#1],\allowbreak \aM[#2],\allowbreak #3,\allowbreak #4)
}}
\newcommand{\SafeR}[2]{\Safe{#2}{#1}}
\newcommand{\SafeFullR}[4]{\SafeFull{#2}{#1}{#4}{#3}}
\newcommand{\SafeDef}{\Safe{\defAg}{\defAgB}}
\newcommand{\SafeDefR}{\SafeR{\defAg}{\defAgB}}
\newcommand{\SafeDefFull}[2]{\SafeFull{\defAg}{\defAgB}{#1}{#2}}
\newcommand{\SafeDefFullR}[2]{\SafeFullR{\defAg}{\defAgB}{#1}{#2}}
\newcommand{\SafeFullDef}{\SafeDefFull{\agTcollVal{}}{\agTcollValB{}}}
\newcommand{\SafeFullDefR}{\SafeDefFullR{\agTcollVal{}}{\agTcollValB{}}}

% \newcommand{\conflictClause}[2]{\ensuremath{\Phi_{\ag[#1], \ag[#2]}}}
% \newcommand{\conflictClauseFull}[4]{\ensuremath{
%   \conflictClause{#1}{#2}(\aM[#1], \aM[#2], #3, #4)
% }}
% \newcommand{\conflictClauseR}[2]{\conflictClause{#2}{#1}}
% \newcommand{\conflictClauseFullR}[4]{\conflictClauseFull{#2}{#1}{#4}{#3}}
% \newcommand{\conflictClauseDef}{\conflictClause{\defAg}{\defAgB}}
% \newcommand{\conflictClauseDefR}{\conflictClauseR{\defAg}{\defAgB}}
% \newcommand{\conflictClauseDefFull}[2]{\conflictClauseFull{\defAg}{\defAgB}{#1}{#2}}
% \newcommand{\conflictClauseDefFullR}[2]{\conflictClauseFullR{\defAg}{\defAgB}{#1}{#2}}
% \newcommand{\conflictClauseFullDef}{\conflictClauseDefFull{\agTcollVar{}}{\agTcollVarB{}}}
% \newcommand{\conflictClauseFullDefR}{\conflictClauseDefFullR{\agTcollVar{}}{\agTcollVarB{}}}

%Instead, these are handled using floating-point simulations externally
%and added to the constraints set
%in the form of conjunction of conflict clauses
%in a~lazy fashion.

Whenever the algorithm $\mathit{findplan}$ from Section~{\ref{sec:alg}} computes a~pre-plan
that still contains a collision,
it represents the generalization of collisions by a~formula~$\collGen$
whose negation it then adds to the formula passed to the SMT solver.
We will now discuss how to first detect collisions
and how to then construct the formula~$\collGen$ generalizing detected collisions.
Here, we will assume precise arithmetic, deferring the discussion of implementation in finite computer arithmetic
to Section~\ref{sec:impl}.

\paragraph{Collision Detection.}

Assume that two agents \ag{} and~\agB{}
follow their motion functions
\(\aM{}: [0,\aD{}] \rightarrow M\)
and \(\aMB{}: [0,\aDB{}] \rightarrow M\) with durations~\aD{} and~\aDB{},
respectively, corresponding to either a~move or a~wait action.
Assume that the agents start the motions
at points in time~\agTcollVal{} and~\agTcollValB{}, respectively.
To determine whether there is a~collision,
we will use the abstract predicate \isCollision{}
introduced in Section~\ref{sec:problem}.
Based on this, we can check for a~collision
of two agents that follow motion functions
starting at certain times:
\begin{definition}\label{def:conflict}
  For two motion functions~\aM{} and~\aMB{}
  with respective starting times~\agTcollVal{} and~\agTcollValB{},
  \Full{\\}%
  \inConflictFullDef{} iff
  \[
  \begin{split}
    \shortTab
    \exists t \in [\agTcollVal{}, \agTcollVal{}+\aD{}] \cap [\agTcollValB{},\agTcollValB{}+\aDB{}]
    \;.\;
    \Short{\\}\shortTab\Short{\qquad}
    \isCollisionFullDef{\aM{}(t-\agTcollVal{})}{\aMB{}(t-\agTcollValB{})}
    .
  \end{split}
  \]
\end{definition}

We will now discuss the construction of the formula~$\collGen$ that generalizes collisions of pre-plans found in Algorithm~\ref{alg:findplan}. A found pre-plan may result in several such collisions. We start with generalizing one of them and consider two cases: The case of a~collision between two moving agents,
and the case of a~collision between a~waiting and a~moving agent.
We can ignore the case when both agents are waiting:
Such a~conflict either should have been avoided
already in the previous discrete steps,
or the agents must in the case of a~conflict overlap
right at the beginning of a~pre-plan,
resulting in a~trivially infeasible plan.

\paragraph{Collisions While Moving.}

In this case, one of the two agents has to wait until the conflict vanishes.
We are interested in waiting the minimal time and hence define
  \(\SafeFullDef{}:=\)
  \[
  \begin{split}
    \inf\{
      \agTcollVar{} \mid\; \shortTab \agTcollVal{} < \agTcollVar{},
      \Short{\\}\shortTab
      \neg \inConflictDefFull{\agTcollVar{}}{\agTcollValB{}}
    \}
    .
  \end{split}
  \]
Note that
\(\agTcollVal{} < \SafeFullDef{} \leq \agTcollValB{}+\aDB{}\). Here,
the lower bound is a consequence of the assumption that agents are open sets which
makes collisions happen in the interior of those sets. The upper bound follows from the fact that \inConflictDefFull{\agTcollValB{}+\aDB{}}{\agTcollValB{}} is always false.

Assume that we detected a~conflict
between two move actions starting at times~\agTcollVal{} and~\agTcollValB{}
and hence \inConflictFullDef{}.
We know that any value of~\agTcollVar{} with \(\agTcollVal{} \leq \agTcollVar{} < \SafeFullDef{}\)
also leads to a~conflict.
In addition---letting the second agent wait---%
any value of~\agTcollVarB{} with \(\agTcollValB{} \leq \agTcollVarB{} < \SafeFullDefR{}\)
also leads to a conflict.

However, we know even more.
For seeing this, observe that \inConflictDef{} is invariant
wrt. translation along the time-axes,
that is,
for every~$\Delta\in\mathbb{R}$,
\inConflictFullDef{} iff
\inConflictDefFull{\agTcollVal{}+\Delta}{\agTcollValB{}+\Delta}
which can be seen by simply translating the witness~$t$ from Definition~\ref{def:conflict} by the same value~$\Delta$.
Due to this, the same conflict happens
for all pairs $(\agTcollVar{}, \agTcollVarB{})$
with the same relative distance as the relative distance
of $(\agTcollVal{}, \agTcollValB{})$.
Hence we know that both
\[
  \agTcollVal{} - \agTcollValB{} \leq \agTcollVar{} - \agTcollVarB{} < \SafeFullDef{} - \agTcollValB{}
\]
and
\[
  \agTcollValB{} - \agTcollVal{} \leq \agTcollVarB{} - \agTcollVar{} < \SafeFullDefR{} - \agTcollVal{}
\]
lead to a~conflict.

Multiplying the second inequality by $-1$, we get
\[
  \agTcollVal{} - \SafeFullDefR{} < \agTcollVar{} - \agTcollVarB{} \leq \agTcollVal{} - \agTcollValB{}
\]
and combining the result with the first inequality, we get
\[
\begin{split}
  \agTcollVal{} - \SafeFullDefR{} < \agTcollVar{} - \agTcollVarB{}
  \\\ \land\
  \agTcollVar{} - \agTcollVarB{} < \SafeFullDef{} - \agTcollValB{}
  .
\end{split}
\]
%whose negation results in one conflict clause
%that captures the information from the two clauses from above:
%\[
%\begin{split}
%  \agTcollVal{} - \SafeFullDefR{} \geq \agTcollVar{} - \agTcollVarB{}
%  \\\ \lor\
%  \agTcollVar{} - \agTcollVarB{} \geq \SafeFullDef{} - \agTcollValB{}
%  ,
%\end{split}
%\]
%which can be rewritten to the final form
%\begin{equation}\label{eq:move:clauses}
%\begin{split}
%  \agTcollVar{} - \agTcollVarB{} \geq \SafeFullDef{} - \agTcollValB{}
%  \\\ \lor\
%  \agTcollVarB{} - \agTcollVar{} \geq \SafeFullDefR{} - \agTcollVal{}
%\end{split}
%\end{equation}
%that is
%\[
%  \conflictClauseFullDef{} \ \land\ \conflictClauseFullDefR{}
%\]
%where
%\begin{equation*}%\label{eq:move:clause}
%\begin{split}
%  &\conflictClauseFullDef{} := \\
%  &\qquad \agTcollVar{} - \agTcollVarB{} <\SafeFullDef{} - \agTcollValB{}
%  ,
%\end{split}
%\end{equation*}
%\begin{equation*}
%\begin{split}
%&\conflictClauseFullDef{} := \\
%  &\qquad \agTcollVar{} - \agTcollVarB{} <\SafeFullDef{} - \agTcollValB{}
%\end{split}
%\end{equation*}

%Recall that~\agTcollVar{} and~\agTcollVarB{}
%are starting times of motions~\aM{} and~\aMB{}
%of agents~\ag{} and~\agB{}, respectively.
%Here we use~\agTcollVal{} and~\agTcollValB{}
%as the values of the arguments~\agTcollVar{} and~\agTcollVarB{}
%when the conflict was detected.

\newcommand{\moveConflictClause}[6]{\ensuremath{
  \varphi^{mm}(\ag{}, \agB{},\allowbreak #1, #2,\allowbreak #3, #4,\allowbreak #5, #6)
}}
\newcommand{\moveConflictClauseDef}{\moveConflictClause{\aM{}}{\aMB{}}{\agTcollVal{}}{\agTcollValB{}}{\agTcollVar{}}{\agTcollVarB{}}}

\newcommand{\agTmoveCollVar}{\ensuremath{\at[\ofAg{j}]{\agTvar{}} + \at[\ofAg{j}]{\agWvar{}}}}
\newcommand{\agTmoveCollVarB}{\ensuremath{\at[\ofAgB{j}]{\agTvarB{}} + \at[\ofAgB{j}]{\agWvarB{}}}}
\newcommand{\agTmoveCollVarVal}{\ensuremath{\at[\ofAg{j}]{\agTval{}} + \at[\ofAg{j}]{\agWval{}}}}
\newcommand{\agTmoveCollVarValB}{\ensuremath{\at[\ofAgB{j}]{\agTvalB{}} + \at[\ofAgB{j}]{\agWvalB{}}}}

For applying this to the variables
of the SMT encoding described in Section~\ref{sec:encoding},
we denote this formula by \moveConflictClauseDef{},
replacing the arguments by the corresponding terms from the SMT encoding.
More concretely, we observe that
the start of a~move action of agent~\ag{} at step~\ofAg{j}
is modeled by the term \agTmoveCollVar{}
and the start of a~move action of agent~\agB{} at step~\ofAgB{j}
by the term \agTmoveCollVarB{}.

Now assume that we detected a~conflict of two agents~\ag{} and~\agB{}
moving along respective edges $(\ofAg{u}, \ofAg{v})$ and $(\ofAgB{u}, \ofAgB{v})$,
starting in discrete steps~\ofAg{j} and~\ofAgB{j}
and times \agTmoveCollVarVal{} and \agTmoveCollVarValB{}
(the hats denoting the values assigned to the respective variables). In this case,
\inConflictDefFull{\agTmoveCollVarVal{}}{\agTmoveCollVarValB{}}, and
the formula~$\collGen$ has the form
\begin{equation*}\label{eq:move:clause:var}
\begin{split}
  &\at[\ofAg{j}]{\agVvar{}} = \ofAg{u} \land \at[\ofAg{j}+1]{\agVvar{}} = \ofAg{v}
  \\\ \land\ &
  \at[\ofAgB{j}]{\agVvarB{}} = \ofAgB{u} \land \at[\ofAgB{j}+1]{\agVvarB{}} = \ofAgB{v}
  \\\ \land\ & \moveConflictClause{\Mof{(\ofAg{u}, \ofAg{v})}}{\Mof{(\ofAgB{u}, \ofAgB{v})}}
    {\\&\qquad\agTmoveCollVarVal{}}{\agTmoveCollVarValB{}}
    {\\&\qquad\agTmoveCollVar{}}{\agTmoveCollVarB{}}.
\end{split}
\end{equation*}

%\begin{align*}\label{eq:move:clause:var}
%  &\at[\ofAg{j}]{\agVvar{}} = \ofAg{u} \land \at[\ofAg{j}+1]{\agVvar{}} = \ofAg{v}
%  \\\ \land\ &
%  \at[\ofAgB{j}]{\agVvarB{}} = \ofAgB{u} \land \at[\ofAgB{j}+1]{\agVvarB{}} = \ofAgB{v}
%  \\\ \land\ &
%  \hat{s}_{a,b} < \at[\ofAg{j}]{\agTvar{}} + \at[\ofAg{j}]{\agWvar{}} - \at[\ofAgB{j}]{\agTvarB{}} - \at[\ofAgB{j}]{\agWvarB{}}
%  \\\ \land\ &
%  \at[\ofAg{j}]{\agTvar{}} + \at[\ofAg{j}]{\agWvar{}} - \at[\ofAgB{j}]{\agTvarB{}} - \at[\ofAgB{j}]{\agWvarB{}} < \hat{s}_{b,a}
%\end{align*}
% &\conflictClauseDefFull{\at[\ofAg{j}]{\agTvar{}} + \at[\ofAg{j}]{\agWvar{}}}{\at[{\ofAgB{j}}]{\agTvarB{}} + \at[{\ofAgB{j}}]{\agWvarB{}}}
% \\\ \land\
% &\conflictClauseDefFullR{\at[\ofAg{j}]{\agTvar{}} + \at[\ofAg{j}]{\agWvar{}}}{\at[{\ofAgB{j}}]{\agTvarB{}} + \at[{\ofAgB{j}}]{\agWvarB{}}}
%  .
%with the constants
%\begin{itemize}
%  \item[] \(\hat{s}_{a,b} = \agTcollVal{} - \SafeFullDefR{}\) and
%  \item[] \(\hat{s}_{b,a} = \SafeFullDef{} - \agTcollValB{}\) where
%  \item[] \(\agTcollVal{} = \agTmoveCollVarVal{}\) and
%  \item[] \(\agTcollValB{} = \agTmoveCollVarValB{}\).
%\end{itemize}

This means that there are~6 possibilities how to resolve such a~conflict
(changing one of the four vertices of edges
along which the two move actions took place
or changing one of the two starting times of the move actions).

%%%%%%% This follows from the fact that G is a graph, not a multi-graph.
%Here we make an~assumption that the graph does not allow
%multiple edges
%with the same starting and ending vertex, that is,
%for all \(e_1, e_2 \in E\), \(e_1 = (u_1,v_1)\), \(e_2 = (u_2,v_2)\),
%\(u_1 = u_2 \land v_1 = v_2\) implies \(e_1 = e_2\).
%Thus, every move action is entirely determined by the enclosing vertices.
%Otherwise it would be necessary to include
%the selected move action into the clause,
%which would not be an~issue because
%the set of move actions is finite.

Now we also discuss conflicts where a~waiting agent participates.

\paragraph{Collisions While Waiting.}

\newcommand{\agX}[1][\defAg]{\ofAg[#1]{x}}
\newcommand{\agTcollPrevVal}[1][\defAg]{\ensuremath{\agTcollVal[#1]}}
\newcommand{\agTcollPrevVar}[1][\defAg]{\ensuremath{\agTcollVar[#1]}}
\newcommand{\agTcollNextVal}[1][\defAg]{\ensuremath{\agTcollVal[#1]'}}
\newcommand{\agTcollNextVar}[1][\defAg]{\ensuremath{\agTcollVar[#1]'}}
\newcommand{\aDprev}[1][\defAg]{\ensuremath{\aD[#1]}}
\newcommand{\aDnext}[1][\defAg]{\ensuremath{\aD[#1]'}}

We also have to ensure that no collisions happen while an~agent~\ag{} is waiting.
In principle, the motion function~\aM{}
can also be constant,
and hence one might be tempted to just specialize the formula
for two moving agents to this case.
However, unlike move actions, wait actions do not have fixed durations, but their duration is a~consequence of the timing of the previous and following move action.
We take this into account, generalizing the given conflict over arbitrarily long wait actions.

So assume an~agent~\ag{} waiting at a~point~$\agX{} \in M$
and an~agent~\agB{} following motion function~\aMB{}
starting from time~\agTcollValB{}.
Assume that a~collision happens at a~certain point in time~$\hat{t}$.
So we have
\(
  \agTcollValB{} \leq \hat{t} \leq \agTcollValB{}+\aDB{}
  \land
  \isCollisionFullDef{\agX{}}{\aMB{}(\hat{t} - \agTcollValB{})}
\).

Let~\agTcollPrevVal{} be the end of the move action of the waiting agent~\ag{} before this waiting period
and let \agTcollNextVal{} be the starting time of the move action
of the waiting agent after this waiting period.
%Let~\aDprev{} and~\aDnext{} the corresponding durations.

\begin{center}
\includegraphics[width=0.9\columnwidth]{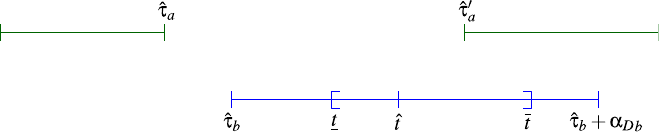}
\end{center}

\newcommand{\waitCollLo}{\ensuremath{\underline{t}_{\:\ag{}, \agB{}, \agX{}, \aMB{}, \agTcollValB{}}}}
\newcommand{\waitCollHi}{\ensuremath{\overline{t}_{\ag{}, \agB{}, \agX{}, \aMB{}, \agTcollValB{}}}}

So we compute the beginning of the collision
\(\waitCollLo{} :=\)
\[
\begin{split}
  \inf \{\agTcollValB{} \leq t \leq \hat{t} \mid
    \isCollisionFullDef{\agX{}}{\aMB{}(t - \agTcollValB{})}
  \}
\end{split}
\]
and its end
\(\waitCollHi{} :=\)
\[
\begin{split}
  % \sup is redefined by some package
  \text{sup} \{\shortTab\hat{t} \leq t \leq \agTcollValB{} + \aDB{} \mid
    \Short{\\}\shortTab\Short{\qquad}
    \isCollisionFullDef{\agX{}}{\aMB{}(t - \agTcollValB{})}
  \}
  .
\end{split}
\]

\newcommand{\waitConflictClause}[6]{\ensuremath{
  \varphi^{wm}(\ag{}, \agB{},\allowbreak #1, #2,\allowbreak #3, #4,\allowbreak #5, #6)
}}
\newcommand{\waitConflictClauseDef}{\waitConflictClause{\agX{}}{\aMB{}}{\agTcollValB{}}{\agTcollPrevVar{}}{\agTcollNextVar{}}{\agTcollVarB{}}}

\newcommand{\agTwaitCollVar}{\ensuremath{\at[\ofAg{j}]{\agTvar{}}}}
\newcommand{\agTwaitEndCollVar}{\ensuremath{\agTmoveCollVar{}}}

So for any wait action of agent~\ag{}
starting at~\agTcollPrevVar{}
and ending at~\agTcollNextVar{}
and any move action of agent~\agB{} starting at~\agTcollVarB{},
the collision happens if
the upper bound~\waitCollHi{} is after the end of the previous action
and the lower bound~\waitCollLo{} is before the beginning of the next action. The result is
\[
\begin{split}
  \agTcollPrevVar{}  - \agTcollVarB{} < \waitCollHi{}& - \agTcollValB{}
  \\\ \land\
  \waitCollLo{}& - \agTcollValB{} < \agTcollNextVar{} - \agTcollVarB{}
  ,
\end{split}
\]
which we will denote by \waitConflictClauseDef{}.
%whose negation we can add as a conflict clause:
%\begin{equation}\label{eq:wait:clause}
%\begin{split}
%  \agTcollVar{} + \aD{} - \agTcollVarB{} &\geq \overline{t} - \agTcollValB{}
%  \\\ \lor\
%  \agTcollNextVar{} - \agTcollVarB{} &\leq \underline{t} - \agTcollValB{}
%  .
%\end{split}
%\end{equation}

%Note that the first row of Equation~\ref{eq:wait:clause}
%is similar to Equation~\ref{eq:move:clause}:
%\(\agTcollVar{} + \aD{}\) here represents the beginning time
%of the wait action,
%and $\overline{t}$ is again a~time
%when the conflict vanishes,
%similarly to \SafeDef{}.

Now we again apply this to the variables of the SMT encoding
described in Section~\ref{sec:encoding},
replacing the arguments of the formula \waitConflictClauseDef{}
by their corresponding terms from the SMT encoding.
So when we detect a~conflict between an~agent~\ag{}
that waits at vertex~\ofAg{u} at time step~\ofAg{j}
and an~agent~\agB{} moving along an edge $(\ofAgB{u}, \ofAgB{v})$
at time step~\ofAgB{j},
starting at \agTmoveCollVarValB{},
the formula~$\collGen$ has the form
\[
\begin{split}
  &\at[\ofAg{j}]{\agVvar{}} = \ofAg{u}
  \\\ \land\
  &\at[\ofAgB{j}]{\agVvarB{}} = \ofAgB{u} \land \at[\ofAgB{j}+1]{\agVvarB{}} =\ofAgB{v}
  \\\ \land\
  & \waitConflictClause{\mathit{coord}(\ofAg{u})}{\Mof{(\ofAgB{u}, \ofAgB{v})}}
    {\\&\qquad\agTmoveCollVarValB{}}
    {\\&\qquad\agTwaitCollVar{}}{\agTwaitEndCollVar{}}
    {\agTmoveCollVarB{}}
  .
\end{split}
\]
%\[
%\begin{split}
%  &\at[\ofAg{j}]{\agVvar{}} = \ofAg{u}
%  \\\ \land\
%  &\at[\ofAgB{j}]{\agVvarB{}} = \ofAgB{u} \land \at[\ofAgB{j}+1]{\agVvarB{}} =\ofAgB{v}
%  \\\ \land\
%  &\at[\ofAg{j}]{\agTvar{}} + \at[\ofAg{j}]{\agWvar{}} +\at[\ofAg{j}]{m} - (\at[\ofAgB{j}]{\agTvarB{}} + \at[\ofAgB{j}]{\agWvarB{}}) %<\overline{t} - (\at[\ofAgB{j}]{\agTcollValB{}} + \hat{\at[\ofAgB{j}]{\agWvarB{}}})
%  \\\ \land\
%  & \underline{t} - (\at[\ofAgB{j}]{\agTcollValB{}} + \hat{\at[\ofAgB{j}]{\agWvarB{}}}) <\at[\ofAg{j}+1]{\agTvar{}} + \at[\ofAg{j}+1]{\agWvar{}} - %(\at[\ofAgB{j}]{\agTvarB{}} + \at[\ofAgB{j}]{\agWvarB{}})  \
%  .
%\end{split}
%\]

We ended up with a~conflict clause
that actually does not depend on a~previous move action.
In the case where the wait action does not have a~next move action,
the conflict clause can be modified in a~straightforward way.

\paragraph{Further Generalization.}

To fully exploit the computational effort that is necessary to find pre-plans, we generate generalizations for all  conflicts a found pre-plan contains. Hence we check for conflicts between all pairs of agents, discrete steps and corresponding move and wait actions.

We further generalize the conflicts
such that we
also check for other conflicts of the given~pair of agents when taking a move action from the same source to a different target vertex.

We did not find it useful
to furthermore generalize the conflicts to
further pairs of agents and/or discrete steps.

%\section{\Uppercase{Theoretical Properties and Comparison}}

\section{\Uppercase{Implementation}}
\label{sec:impl}

\newcommand{\agC}[1][\defAg]{\ofAg[#1]{c}}
\newcommand{\agCB}{\agC[\defAgB]}

\paragraph{Collision Detection and Avoidance.}

We implemented the predicates and functions introduced in Section~\ref{sec:coll} as follows:
\begin{itemize}
  \item We assume that each agent~\ag{} is abstracted into an~open disk
  with a~fixed radius~\(\agR{} \in \mathbb{R}^{>0}\).
  Hence for agents~\ag{} and~\agB{},
  \isCollisionFullDef{\agC{}}{\agCB{}} iff
  \(||\agC{}-\agCB{}|| < \agR{}+\agRB{}\),
  where \( \agC{}, \agCB{} \in M \)
  are respective centers of the disks of the agents.
  \item We assume that the agents move with constant velocity
  following straight lines of the edges.
  As a~result, \inConflictDef{} corresponds to checking
  whether a~quadratic inequality has a~solution
  in the intersection of the time intervals from Definition~\ref{def:conflict}.
  \item Exploiting the observation that
  \(\SafeFullDef{} \in (\agTcollVal{}, \agTcollValB{}+\aDB{}]\)
  we compute the resulting value by
  binary search of the switching point~\agTcollVar{}
  in the interval
  for which $\neg\inConflictDefFull{\agTcollVar{}}{\agTcollValB{}}$
  and $\inConflictDefFull{\agTcollVar{} - \epsilon}{\agTcollValB{}}$,
  for a~small enough \(\epsilon > 0\).
\end{itemize}

\paragraph{Floating-point Numbers.}
Our simulations of collisions of agents
are based on floating-point computation whereas SMT solvers treat linear real arithmetic precisely, using rational numbers for all computation.
There are two main issues here:
\begin{itemize}
  \item Conversion of a floating point number to a rational number may result in huge integer values
  for the numerator and denominator, although the intended value
  is very close to a~simple rational or even integer number.
  \item Conversion of rational numbers to floating point numbers, and the following computation in floating point representation may incur approximation errors (e.g., due to round-off or discretization). For example, this may lead to the situation where---in the case of a collision between two moving agents---the added conflict clause does not require the waiting agent to wait long enough to completely avoid the same collision. Hence a~very close collision may re-appear, and the same situation may repeat itself several times.

  %resulting in
  %can result in a~value that is very close to the original floating-point value
  %but still different.
  %Consider the following simplified example:
  %simulations produced floating-value~1.777
  %which is a~lower bound to avoid a~collision.
  %The resulting rational number may be e.g.~$\frac{227}{128}$,
  %which however corresponds to a~floating-point value close to~1.773.
  %Thus, the SMT solver can try to resolve the conflict
  %by setting the arrival time of the corresponding agent to~$\frac{227}{128}$,
  %which will, however, again result in a~collision
  %in the floating-point-based simulation.
\end{itemize}

We overcome these deficiencies with an~overapproximation
of the conflict intervals
along with simplification of the resulting rational numbers
using simple continued fractions and best rational approximation:
%% I just could not find a paper on this except of Wikipedia
in the case that a~floating-point value~$x$
is respectively a~lower or a~higher bound of a~conflicting interval,
the result corresponds to best rational approximation
from $(x-\epsilon, x]$ or $[x, x+\epsilon)$, respectively,
for an~$\epsilon$ that is large enough.
This not only avoids the re-appearance of the same conflict, but also maps
floating-point values that are close to each other to the same rational numbers, avoiding the appearance of tiny differences between rational numbers that tend to clog the SMT solver.

\paragraph{Heuristics.}
The formula passed to the SMT solver often represents
a~highly underconstrained problem,
spanning a~huge solution space.
Due to this, it is essential that the SMT solver chooses
a~solution in a~goal oriented way
in order to maximize the chances of hitting upon a~$\delta$-optimal plan,
or at least to concentrate search on the most promising part
of the solution space which also concentrates
the addition of conflict avoidance clauses to this part.
%We encode the whole graph into the formula,
%that is,
%we include all possible transitions between vertices
%that correspond to variables
%\at{\agVvar{}} and \at[j+1]{\agVvar{}}.
%Therefore,
%it is very important to guide the search
%since usually only a~small subset of all the corresponding constraints
%is necessary to explore.
For this we prefer transitions to vertices
that lie on shorter paths to the goal
over transitions to vertices that lie on longer paths.
This can be easily precomputed using Dijkstra's algorithm
for all vertices with a~fixed start and goal.

Nonetheless, using such heuristics
does not evade the problem
of encoding all the transitions into the formula,
which floods the SMT solver with a~lot of constraints
that are not essential at arriving at the desired plan.
Also, conversion of the resulting formula to CNF might be expensive.

\paragraph{Tools.}
We implemented the resulting algorithm on top of MathSAT5~\cite{Cimatti:13} SMT solver.
We incrementally build the formula described in Section~\ref{sec:encoding}
using API.
However, since we do not even require the SMT solver itself
to handle optimization,
it is easy to replace the API calls to another SMT solver
that handles \LRA{}.
We also implemented a~visualization tool of \MAPFR{} problems.
Our tools are available online and are open-source.

\section{\Uppercase{Computational Experiments}}
\label{sec:exp}

We compare run-times
of our implementation from Section~\ref{sec:impl}
denoted as \smtLra{}
and state-of-the-art tools
\ccbs{} and \smtCcbs{},
both presented in~\cite{AndreychukYSAS22},
which define the \MAPFR{} problem in a~similar way.
These tools search for optimal plans,
which is more difficult than
searching for sub-optimal plans,
such as minstep $\delta$-optimal plans in our case.
However, as discussed in the introduction,
not only that the price for getting optimal plans may be too high,
but the resulting plans may also be not an~ideal fit in practice,
due to possible requirements on flexible dispatchability of the plans,
and due to the fact that the dynamics of the agents
may not be modeled accurately.
Based on these observations,
we consider the comparisons to be practically reasonable.
%+ actual distance to the optimum

There are also differences concerning
the function that is being optimized
which is
sum of costs in the case of \ccbs{}
and makespan in the case of \smtCcbs{}.
We support both of these cost functions in the form of a~parameter.
While there are certainly instances where the choice of the cost function
qualitatively matters,
\cite{AndreychukYSAS22} showed that
both the tools yield similar respective costs
of the resulting plans within their benchmarks.
Hence, comparing such tools with different objectives also makes sense.

In the following experiments,
we will use a~similar setup to~\cite{AndreychukYSAS22},
that is,
similar to both of the presented state-of-the-art tools.

We will start by
the description of the benchmarks.
Finally, we will present and discuss computational results
of the performed experiments.

\subsection{Description of Benchmarks}

\newcommand{\bench}[1]{\texttt{#1}}
\newcommand{\emptyB}{\bench{empty}}
\newcommand{\roadB}{\bench{road\-map}}
\newcommand{\bottleB}{\bench{bottle\-neck}}

A~benchmark is specified by a~graph
and a~set of agents,
each defined by a~radius
and
a~starting and goal vertex.
The following benchmarks use the same radius for all agents,
and hence we will not discuss radii any more.

We did experiments with three classes of problems:
\emptyB{}, \roadB{} and \bottleB{}.
Benchmarks \emptyB{} and \roadB{}
are adopted from~\cite{AndreychukYSAS22}
and correspond to MAPF maps from the Moving AI repository.
Our \bottleB{} benchmark is an~additional simple experiment
which identifies a~weakness of the state-of-the-art approaches.
More detailed description of the benchmarks follows below.

\New{\Short{
However,
we discuss benchmark \roadB{}
only in the extended version of the paper\footnote{\url{https://arxiv.org/abs/2312.08051}},
along with further figures, plots and tables.
}}

We did not include benchmarks with large graphs
(i.e. with high number of vertices or edges),
because our current algorithm
encodes the whole graph into the formula,
as discussed within heuristics in Section~\ref{sec:impl}.

\paragraph{Empty Room.}

This benchmark is based on  a~graph that represents an~empty square room
with $16 \times 16$ vertices---the result of grid approximations of MAPF maps
from the Moving AI repository~\cite{AndreychukYSAS22}.

Interconnection of the vertices depend
on a~neighborhood parameter~$n$,
which defines that each vertex
has exactly $2^n$ neighbors
(with the exception of boundary vertices).
For example,
\(n=2\) corresponds to square grid,
\(n=3\) extends the square grid of diagonal edges, etc.
Using such a~graph,
it may be necessary to include a~high amount of useless edges
in order to cover a~suitable number of realistic movements of the agents.
On the other hand,
it might be possible to exploit the fact
that such graphs are highly symmetric.

The resulting benchmark \emptyB{}
represents a~model with no physical obstacles.
Still, with an~increasing number of agents~$k$,
the number of possible collisions of the agents grows significantly,
because most of the shortest paths lead via central regions of the graph.

\Full{
\paragraph{Roadmap.}

Unlike the previous benchmark, here the maps from the Moving AI repository
are not approximated based on grids
but based on the ``roadmap-generation tool
from the Open Motion Planning Library (OMPL),
which is a~widely used tool in the robotics community''.
Such an~approximation results in asymmetric graphs
with possibly very different lengths of edges.
On one hand, such edges can model realistic route choices of the agents.
On the other hand, the number of possible places
where agents are allowed to wait
in order to avoid collisions may be too low,
if the edges are too long---since we only allow waiting at vertices.

We follow the original benchmarks
where the roadmap-generation was applied
on a~large map \bench{den520d}
which comes from the field of video-games.
It is possible to set various levels of discretization (i.e. density)
of the original map.
Here, we only experimented
with benchmarks with the lowest density,
denoted as \bench{sparse}.
}

\paragraph{Bottleneck.}

Benchmark \bottleB{} models the problem
of steering~$k$ agents from~$k$ initial vertices
through a~single transfer vertex to~$k$ goal vertices.
Hence the transfer vertex represents a~bottleneck
every agent has to pass through.
We place the initial and goal vertices
(i.e., altogether $2k$ vertices)
on a~circle whose center is formed by the transfer vertex.
\Full{%
An~example of the benchmark with \(k = 4\) is illustrated
in the Figure~\ref{fig:bottleneck},
where the bigger colored disks denote the agents at their starting positions
and the smaller disks denote the vertices of the graph,
where the colored ones in addition indicate goal positions of the corresponding agents.

\begin{figure}[!htb]
\begin{center}
\includegraphics[trim={10cm 1cm 10cm 1cm},clip,width=.6\columnwidth]{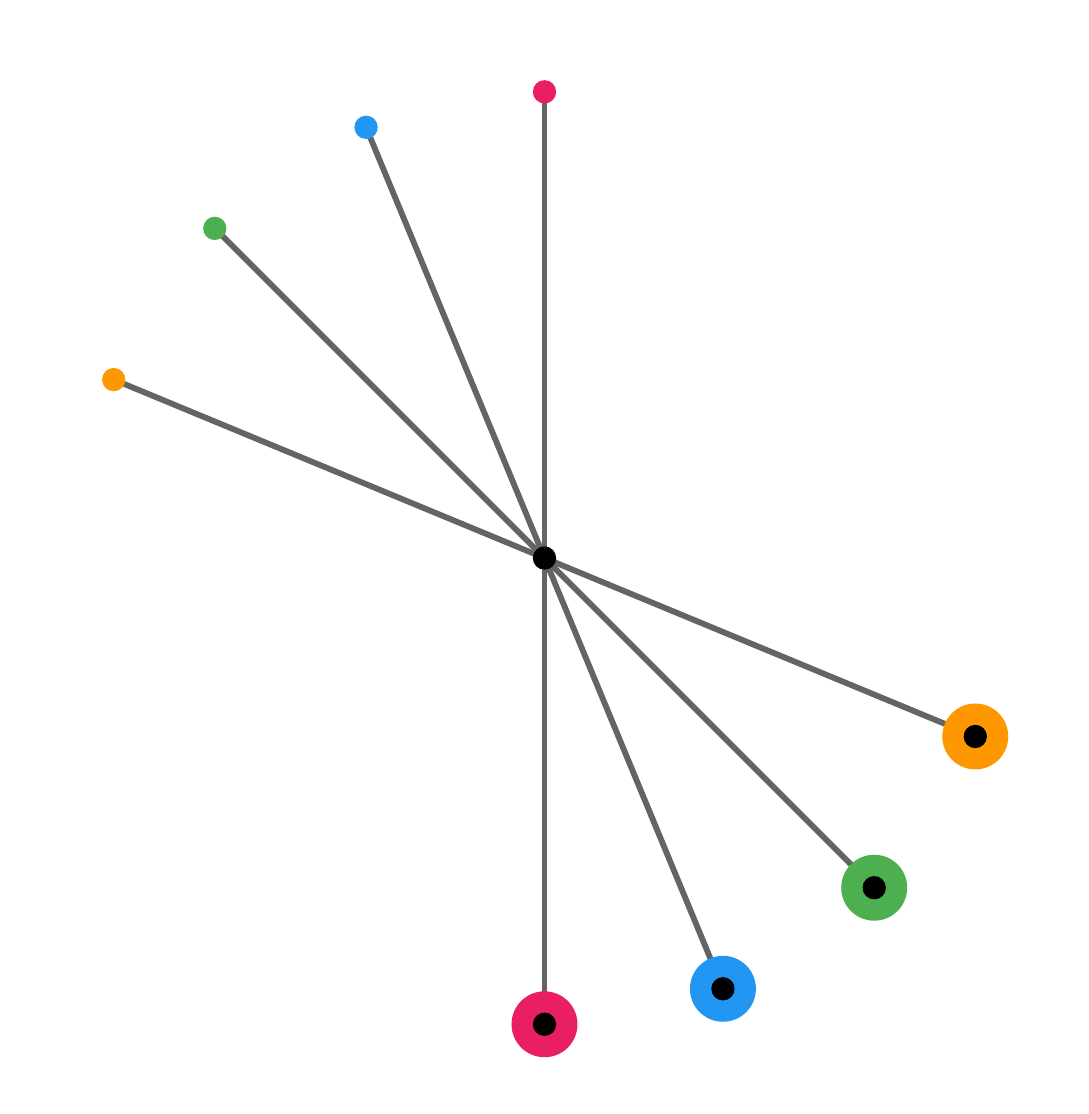}
\end{center}
\caption{Bottleneck benchmark with $k = 4$ agents.}
\label{fig:bottleneck}
\end{figure}
}

The task here boils down to just choosing a~certain order of the agents.
Resolving such a~benchmark problem
can still result in an~exponential complexity
in the number of agents~$k$---%
if just various permutations of the agents are tried,
without a~thorough exclusion of the conflicting time intervals
of particular agents.

\subsection{Experimental Setup}

\newcommand{\lraMakespan}{\texttt{M}}
\newcommand{\lraSOC}{\texttt{S}}

In the case of
\Full[benchmark \emptyB{}]{benchmarks \emptyB{} and \roadB{}},
we observe whether particular experiments finish within a~given timeout.
The set of experiments contains instances where the number of agents ranges from~1 to~64
(none of the tools managed to finish with more agents within the selected timeouts).
For each number of agents~$k\in \{ 1,\dots, 64\}$,
each start and goal vertex of each agent is pre-generated
in 25 random variants.
Here, when generating the variants for agent $k+1$,
all the previous~$k$ agents are reused
and only the positions of the new one are generated randomly.
The result is $64\times 25=1600$ instances
\Full{for each of \emptyB{} (}%
for each neighborhood~$n$\Full{) and \roadB{}}.

The subject of our interest is
how the evaluated algorithms scale with time,
so we ran all the experiments with different timeouts
ranging from~30 seconds up to~16 minutes
(with exponential growth)
and observed how many instances finished in time.
We will show the results in the form of box plots.

\Full{
To also directly illustrate the relationship
of the number of solved instances and~$k$,
we will also show success rate plots,
that is,
plots with the ratios of the number of solved instances
out of the total number of instances (i.e. out of~25)
wrt. a~given number of agents~$k$,
and with a~fixed timeout.
}

Our tool \smtLra{} is in addition parametrized by a~cost function---%
either makespan or sum of costs---%
and by a~sub-optimal coefficient \(\delta \in \{1, \frac{1}{2}, \frac{1}{4}\}\).
In the plots, the parameters are denoted
in the form $(C,\delta)$,
where~$C$ is either~\lraMakespan{} (makespan)
or~\lraSOC{} (sum of costs).
In all experiments,
the higher~$\delta$ was, the more instances were solved.
Hence, to make the box plots more compact,
we merged all the variants of~$\delta$
related to the same cost function
such that the boxes of the variants with lower~$\delta$
overlay the boxes of the variants with higher~$\delta$.
Also, higher values of~$\delta$
correspond to lighter colors.
For example,
boxes \(\delta = \frac{1}{4}\) overlay
boxes \(\delta = \frac{1}{2}\),
but the magnitudes of both boxes are still visible
since the number of solved instances
is always lower in the case of \(\delta = \frac{1}{4}\)
than in the case of \(\delta = \frac{1}{2}\).
As a~result,
for each timeout in the box plot,
our tool always takes two columns,
each consisting of three (overlaid) boxes.
\Full{%
In the case of success rate plots,
we use dashed curves in the case of our tool
in order to increase readability,
and include all curves that correspond to the variants of~$\delta$,
where again higher values of~$\delta$
correspond to lighter colors.
}

\Full{
We will provide tables to further illustrate the effect of parameter~$\delta$.
For this, observe that Algorithm~\ref{alg:main}
terminates as soon as
$\Cost{p_{\mathit{opt}}}\leq (1+\delta) t_{\mathit{min}}$,
which ensures $\delta$-optimality.
However, the ratio $\Cost{p_{\mathit{opt}}} / t_{\mathit{min}}$,
that we call \emph{guaranteed ratio},
may actually be well below the required value $1+\delta$,
meaning that the algorithm produced a~better plan than required.
The tables contain the average of the guaranteed ratio of plans
that finished within 16~min
(with lower timeouts, the ratios are even lower).
}

In the case of benchmark \bottleB{},
we only focus on runtime of the evaluated tools
for some numbers of agents ranging from~2 to~30.
In the case of our implementation,
we again include all the variants of parameters mentioned above.
We used timeout 30~min to set some upper boundary
on the runtimes of the tools.

We executed all the benchmarks
on a~machine with Intel(R) Xeon(R) Gold 6254 CPU @ 3.10GHz,
with 180 cores and 1TB memory.
To unify the runtime environment,
we reused and adapted the scripts from the previous experiments~\cite{AndreychukYSAS22}
which are a~part of the \smtCcbs{} tool.
These scripts do not exploit all the available resources of the machine, though.
Still, none of the evaluated tools use parallel computation---%
the cores are only used to run multiple benchmarks concurrently.

\subsection{Results}

\paragraph{Empty Room.}

Recall that benchmark \emptyB{} is parametrized by its neighborhood~$n$
which means that vertices have approximately~$2^n$ neighbors.
We did experiments with $n \in \{2,3,4,5\}$,
all of which are shown in box plots in
Figure~\ref{fig:empty:solved}.
\Full{%
We also provide Table~\ref{tab:empty:coef}
with guaranteed ratios of the resulting plans,
depending on~$n$.
}

\newcommand*{\subfigurewidthratio}{0.49}
\newcommand*{\subfiguregraphicswidthratio}{1.1}
\newcommand*{\singlesubfigurewidthratio}{0.55}

\begin{figure*}[!htb]
\begin{center}
  \begin{subfigure}{\subfigurewidthratio\textwidth}
    \caption{$n = 2$}
    \label{fig:empty:solved:n2}
    \includegraphics[trim={0cm 0.5cm 0cm 0cm},width=\subfiguregraphicswidthratio\columnwidth]{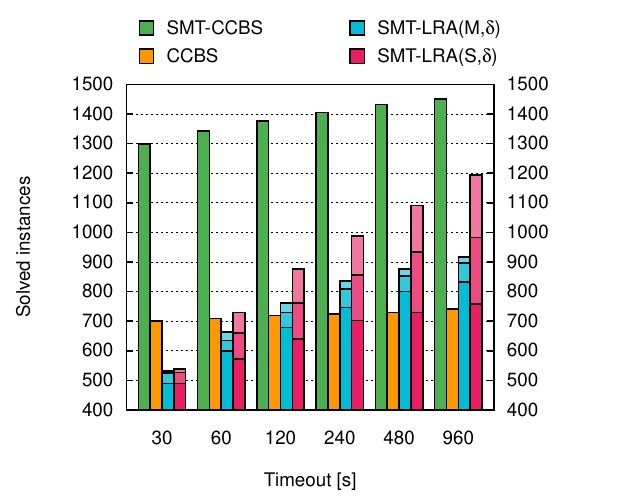}
  \end{subfigure}
  \hfill
  \begin{subfigure}{\subfigurewidthratio\textwidth}
    \caption{$n = 3$}
    \label{fig:empty:solved:n3}
    \includegraphics[trim={0cm 0.5cm 0cm 0cm},width=\subfiguregraphicswidthratio\columnwidth]{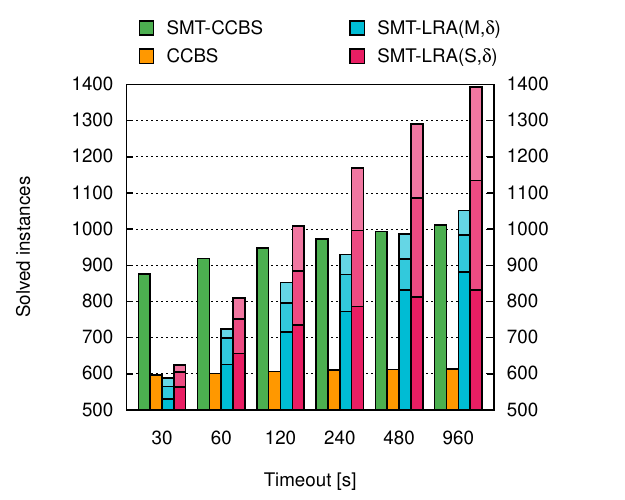}
  \end{subfigure}
\end{center}
\begin{center}
  \begin{subfigure}{\subfigurewidthratio\textwidth}
    \caption{$n = 4$}
    \label{fig:empty:solved:n4}
    \includegraphics[trim={0cm 0.5cm 0cm 0cm},width=\subfiguregraphicswidthratio\columnwidth]{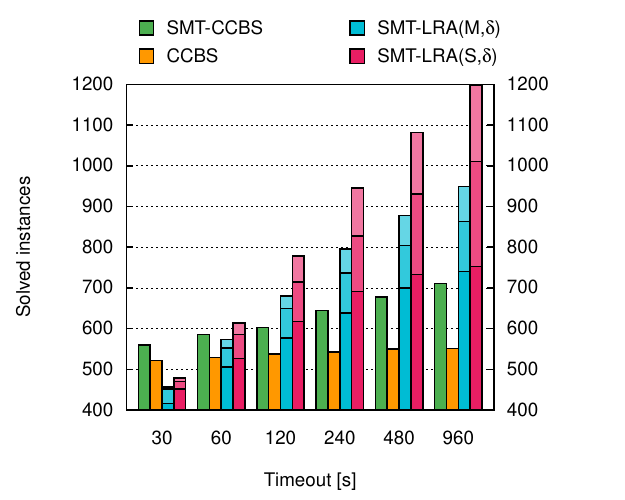}
  \end{subfigure}
  \hfill
  \begin{subfigure}{\subfigurewidthratio\textwidth}
    \caption{$n = 5$}
    \label{fig:empty:solved:n5}
    \includegraphics[trim={0cm 0.5cm 0cm 0cm},width=\subfiguregraphicswidthratio\columnwidth]{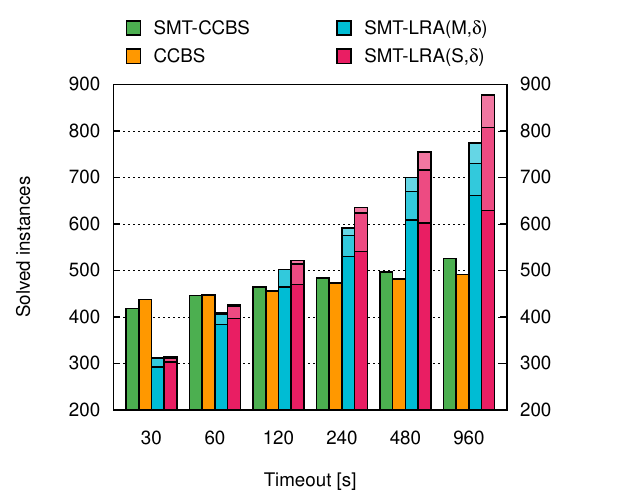}
  \end{subfigure}
\end{center}
\caption{Comparison of the number of solved instances of benchmark \emptyB{} with given~$n$ wrt. a~given timeout.}
\label{fig:empty:solved}
\end{figure*}

\Full{
\begin{table}[!htb]
\caption{Average guaranteed ratios of \smtLra{} in benchmark \emptyB{} wrt. the parameters of the tool and a~given $n$.}
\label{tab:empty:coef}
\begin{center}
\small
\begin{tabular}{|l||c|c|c|c|}
  \hline
  $(C, \delta)\ \text{\textbackslash{}}\ n$ & 2 & 3 & 4 & 5
  \\\hline\hline
  $(\lraMakespan{}, 1   )$ & 1.45 & 1.55 & 1.54 & 1.43  \\\hline
  $(\lraSOC{},      1   )$ & 1.56 & 1.57 & 1.56 & 1.40  \\\hline\hline
  $(\lraMakespan{}, 0.5 )$ & 1.29 & 1.33 & 1.32 & 1.29  \\\hline
  $(\lraSOC{},      0.5 )$ & 1.30 & 1.30 & 1.31 & 1.26  \\\hline\hline
  $(\lraMakespan{}, 0.25)$ & 1.16 & 1.18 & 1.18 & 1.17  \\\hline
  $(\lraSOC{},      0.25)$ & 1.16 & 1.15 & 1.15 & 1.14  \\\hline
\end{tabular}
\end{center}
\end{table}
}

We firstly focus on the comparison of the selected cost functions
in the case of our tool.
We see that usually the cases that optimize sum of costs
perform better than the cases optimizing makespan.
Observe, though, that the results are similar
in the case of \(\delta = \frac{1}{4}\)
which correspond to the boxes at the base.
We explain these as that in the case of sum of costs
there are more possibilities how to reduce the cost of the plan
than in the case of makespan
where the cost usually depends on just one agent,
regarding the symmetry of the graph.
We assume that at the same time this is the reason
why, in the case of makespan,
there are lower increases of the number of solved instances
with growing~$\delta$ compared to the variant
which optimizes sum of costs.
Also notice that in the cases of neighborhood~\(n=2\)
and especially \(n=5\),
there are quite low performance growths
with increasing~$\delta$\Full[. \New{However}]{,
which may be caused by the fact
that the actual guaranteed ratio is lower than in other cases of~$n$,
namely with~\(\delta = 1\)
(see Table~\ref{tab:empty:coef}).
Furthermore},
\cite{AndreychukYSAS22} showed
that these corner cases of~$n$
are actually the least useful benchmarks:
benchmarks with \(n=3\)
offer much faster plans than in the case of \(n=2\),
and \(n=5\) on the other hand
provide only very low improvement over the case of \(n=4\).
All in all, our approach scales well with the growing timeouts,
in every case of neighborhood~$n$,
cost function and parameter~$\delta$.

Now we also focus on the state-of-the-art tools,
where we will actually confirm the observations
made in~\cite{AndreychukYSAS22}.
These tools are consistent in the sense
that the lower parameter~$n$,
the faster is their algorithm---%
because there are less possible paths to the goals.
In our case, the observation holds as well,
but with one exception
in the case of \(n=2\) vs. \(n=3\),
where the runtime of the experiments with the lowest neighborhood is higher.
The reason is that the graphs with higher~$n$
allow that the shortest paths to the goals
take less edges---%
which in our case becomes more important
than the number of possible choices,
because our current algorithm is sensitive
to the number of edges in the graph
which we all encode into the formula.

The state-of-the-art tools usually perform better than our tool
when the timeout is less or equal to 1~min.
\smtCcbs{} performs especially well
in the case of \(n=2\)
because it maps
a~lot of time points to the same values
since many of them are integer values.
We however consider this case to be the least
useful benchmark
referring the earlier discussion
and in addition since the square grids
are not too realistic models
and can also be handled by standard MAPF approaches
(which are currently much faster than \MAPFR{} approaches).
Although \smtCcbs{} scales better with time than \ccbs{},
the highest growth of the number of solved instances
still occurs in the case of our approach,
even in the cases of \(\delta = \frac{1}{4}\)
which correspond to the boxes of our algorithm at the base.
In the cases of \(n \geq 3\) and \(\delta = 1\),
which correspond to the upmost boxes of our algorithm,
we outperform the state-of-the-art tools if the timeout is high.

\Full{
We also provide Figure~\ref{fig:empty:success_rate:t480}
with success rate plots.
Here we selected timeout 8~min (480~s).
Recall that
for our tool we distinguish the curves
that correspond to higher~$\delta$ by lighter colors.
Within a~single cost function,
especially in the case of sum of costs,
the distances between the curves of particular cases of~$\delta$
seem to be quite uniform,
which confirms the observations based on Figure~\ref{fig:empty:solved}
that in many cases our algorithm scales well with the parameter~$\delta$.
At the same time,
the plots also confirm that sometimes in the case of makespan
the performance does not increase much with growing~$\delta$.

\begin{figure*}[!htb]
\begin{center}
  \begin{subfigure}{\subfigurewidthratio\textwidth}
    \caption{$n = 2$}
    \label{fig:empty:success_rate:t480:n2}
    \includegraphics[trim={0cm 0.5cm 0cm 0cm},width=\subfiguregraphicswidthratio\columnwidth]{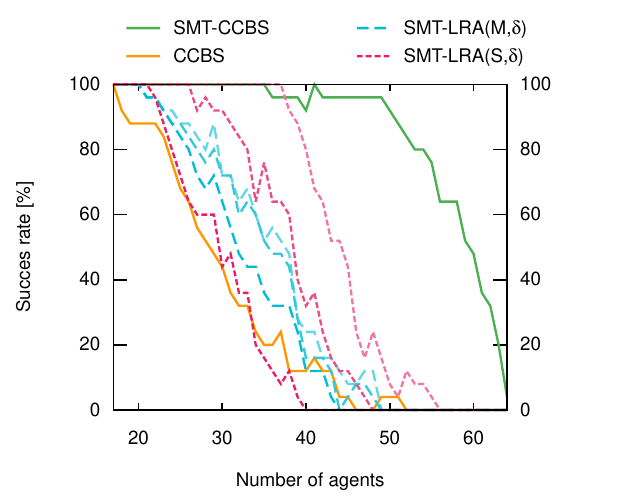}
  \end{subfigure}
  \hfill
  \begin{subfigure}{\subfigurewidthratio\textwidth}
    \caption{$n = 3$}
    \label{fig:empty:success_rate:t480:n3}
    \includegraphics[trim={0cm 0.5cm 0cm 0cm},width=\subfiguregraphicswidthratio\columnwidth]{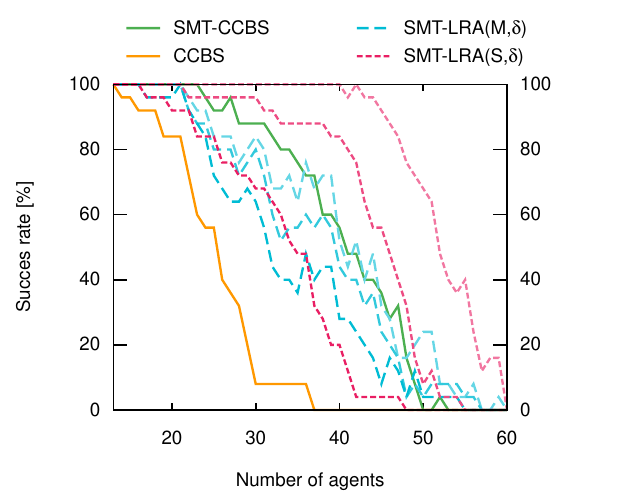}
  \end{subfigure}
\end{center}
\begin{center}
  \begin{subfigure}{\subfigurewidthratio\textwidth}
    \caption{$n = 4$}
    \label{fig:empty:success_rate:t480:n4}
    \includegraphics[trim={0cm 0.5cm 0cm 0cm},width=\subfiguregraphicswidthratio\columnwidth]{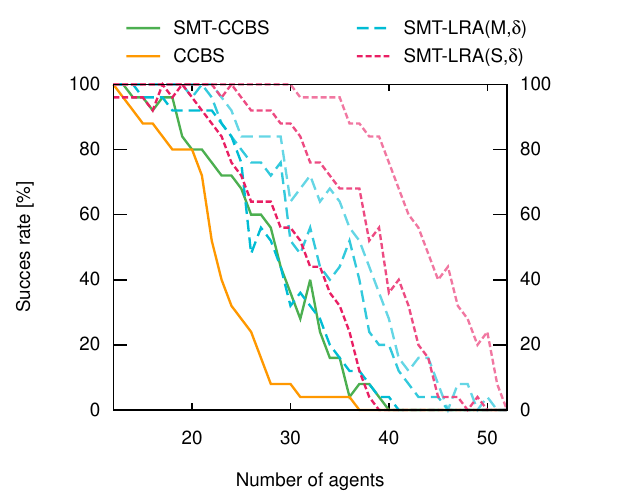}
  \end{subfigure}
  \hfill
  \begin{subfigure}{\subfigurewidthratio\textwidth}
    \caption{$n = 5$}
    \label{fig:empty:success_rate:t480:n5}
    \includegraphics[trim={0cm 0.5cm 0cm 0cm},width=\subfiguregraphicswidthratio\columnwidth]{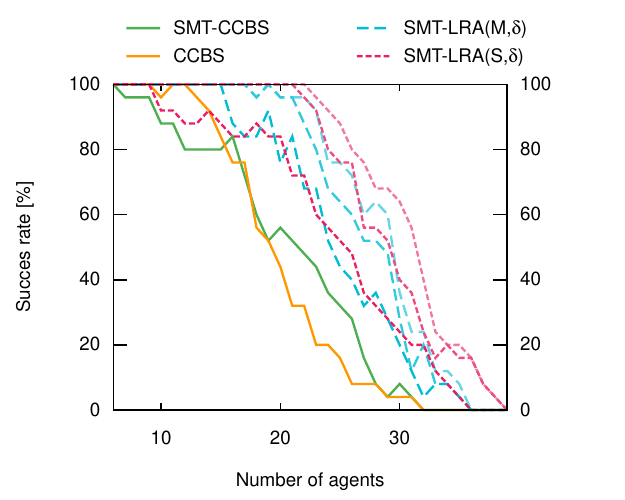}
  \end{subfigure}
\end{center}
\caption{Comparison of the success rates of benchmark \emptyB{} with given~$n$ within timeout 8~min wrt. a~number of agents.}
\label{fig:empty:success_rate:t480}
\end{figure*}

In the case of all the tools,
especially in the case of our tool,
there happen to be glitches in the success rates---%
sometimes the performance increases a~bit with a~higher number of agents.
In some cases, it is probably just caused by inaccurate measurement,
however approaches that are based on a~SAT solver (\smtCcbs{})
or even an~SMT solver (\smtLra{})
may naturally exhibit such behavior
since the algorithms are more complex and not that straightforward
like CBS-based algorithms.
}

\Full{
\paragraph{Roadmap.}

\begin{figure}[!htb]
\begin{center}
\includegraphics[width=\singlesubfigurewidthratio\textwidth]{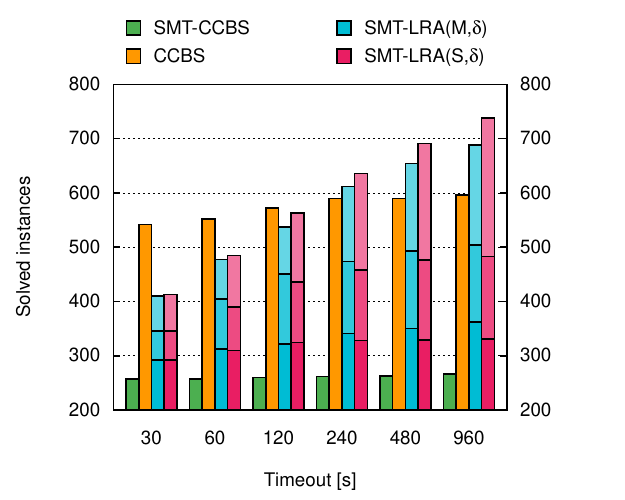}
\end{center}
\caption{Comparison of the number of solved instances of benchmark \roadB{} wrt. a~given timeout.}
\label{fig:roadmap:solved}
\end{figure}

Figure~\ref{fig:roadmap:solved}
shows how particular tools scale with time
in the case of benchmark \roadB{}.
%! different results than in \cite{AndreychukYSAS22}
Clearly, \smtCcbs{} does not handle this benchmark well,
independent of the chosen timeout.
By contrast, \ccbs{} performs very well, especially with smaller timeouts.
However, at the timeout of 4~min (240~s) it reaches a~point
after which it almost stops scaling with time at all.
In our case, the variants which optimize sum of costs or makespan
perform almost the same.
However, the performance depends a~lot on parameter~$\delta$.
For example, with \(\delta = 1\),
which corresponds to the upmost boxes,
our approach scales very well and outperforms \ccbs{}
for timeouts greater or equal to 4~min.
However, in the cases of \(\delta = \frac{1}{4}\)
(i.e., the boxes at the base),
the algorithm is not competitive with \ccbs{} and scales poorly.
We explain this as follows:
the \roadB{} graph is highly asymmetric and contains a~lot of long edges,
compared to the graphs in benchmark \emptyB{}.
Therefore, the shortest paths to the goals
often consist of a~low number of edges.
At the same time,
paths to the goals with similar distances
can actually consist of different number of edges.
Thus, once we find a~(collision-free) plan
and fix the number of steps~$h$ for all agents,
it may happen that when optimizing the plan,
we miss alternative paths that consist of more steps
which could be essential to arriving at easier possibilities
of finding faster plans.

In addition, we provide Table~\ref{tab:roadmap:coef}
with the guaranteed ratios of our tool.
The ratios are quite high in the cases of \(\delta = 1\),
which can explain why the difference of the number of solved instances
is so high compared to \(\delta = \frac{1}{2}\),
and also compared to benchmark \emptyB{},
where the guaranteed ratios are lower.

\begin{table}[!htb]
\caption{Average guaranteed ratios of \smtLra{} in benchmark \roadB{} wrt. the parameters of the tool.}
\label{tab:roadmap:coef}
\begin{center}
\small
\begin{tabular}{|l||c|}
  \hline
  $(C, \delta)$ &
  \\\hline\hline
  $(\lraMakespan{}, 1   )$ & 1.64  \\\hline
  $(\lraSOC{},      1   )$ & 1.65  \\\hline\hline
  $(\lraMakespan{}, 0.5 )$ & 1.31  \\\hline
  $(\lraSOC{},      0.5 )$ & 1.31  \\\hline\hline
  $(\lraMakespan{}, 0.25)$ & 1.14  \\\hline
  $(\lraSOC{},      0.25)$ & 1.14  \\\hline
\end{tabular}
\end{center}
\end{table}

Similarly to benchmark \emptyB{},
we also provide a~plot with success rates of the tools,
in Figure~\ref{fig:roadmap:success_rate:t480},
again with the timeout of 8~min.
Here the success rates are well distributed with no anomalies.

\begin{figure}[!htb]
\begin{center}
\includegraphics[width=\singlesubfigurewidthratio\textwidth]{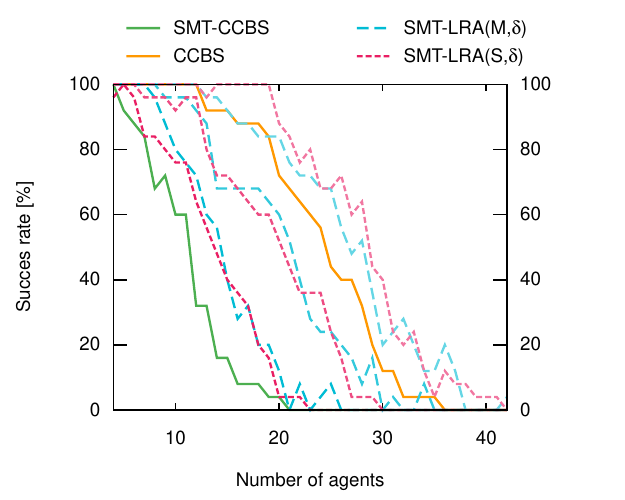}
\end{center}
\caption{Comparison of the success rates of benchmark \roadB{} within timeout 8~min wrt. a~number of agents.}
\label{fig:roadmap:success_rate:t480}
\end{figure}
}

%+ distance to optimum
%+ number of non-optimally solved instances - can be included in plots, but would req. additional text

\paragraph{Bottleneck.}

We summarize the runtimes
of benchmark \bottleB{} of particular tools
in Table~\ref{tab:bottleneck}.
In the case of \smtCcbs{},
we excluded the built-in verification of the solutions
which here seemed to be very time-consuming.
In our case we merge all the corresponding cases of parameter~$\delta$
into single column since the runtimes were almost the same
regardless the parameter.
We further merge both cost functions~\lraMakespan{} and~\lraSOC{}
into one column
s.t. the respective runtimes are separated by a~pipe character.
\Full{%
The average guaranteed ratio of our presented plans
is~1.08 in the case of makespan (\lraMakespan{})
and~1.04 in the case of sum of costs (\lraSOC{}).
}

\begin{table}[!htb]
\caption{Comparison of the runtimes in seconds of benchmark \bottleB{} with a~given number of agents.}
\label{tab:bottleneck}
\begin{center}
\small
\begin{tabular}{|r||c|c|c|c|}
  \hline
  $k$ & \smtCcbs{} & \ccbs{} & \smtLra{}$(\lraMakespan{} | \lraSOC{}\!)$
  \\\hline\hline
  2  & 0.00     & \quad 0.00 & 0.01 $|$ 0.01  \\\hline
  3  & 0.47     & \quad 0.00 & 0.02 $|$ 0.02  \\\hline
  4  & $> 1800$ & \quad 0.00 & 0.04 $|$ 0.02  \\\hline
  5  & ?        & \quad 0.01 & 0.03 $|$ 0.03  \\\hline
  6  & ?        & \quad 0.05 & 0.07 $|$ 0.05  \\\hline
  7  & ?        & \quad 0.45 & 0.08 $|$ 0.06  \\\hline
  8  & ?        & \quad 3.52 & 0.15 $|$ 0.08  \\\hline
  9  & ?        & \ \ 43.99  & 0.15 $|$ 0.11  \\\hline
  10 & ?        & 720.27     & 0.22 $|$ 0.14  \\\hline
  11 & ?        & $> 1800$   & 0.26 $|$ 0.17  \\\hline \hline
  15 & ?        & ?          & 0.46 $|$ 0.42  \\\hline
  20 & ?        & ?          & 0.79 $|$ 0.97  \\\hline
  30 & ?        & ?          & 4.88 $|$ 5.63  \\\hline
\end{tabular}
\end{center}
\end{table}

It is clear that runtimes of both state-of-the-art solvers
exhibit exponential relationship with the number of agents~$k$,
while our algorithm is much less sensitive.
For example, \ccbs{} is fastest until \(k=6\)
but after that point our \smtLra{} dominates the runtime.
The reason is that we resolve the conflicts of agents
using the learning mechanism of generalized conflict clauses
where the timing constraints
efficiently exclude inappropriate orderings of the agents,
making the benchmark fairly easy for our approach---%
which is consistent with the observation
that such a~problem is indeed trivial,
as discussed in the description of benchmarks.
For example, the problem is easily solvable
using an~ad-hoc approach.
Nevertheless,
such bottlenecks may appear as a~part of more complex problems
where a~sophisticated algorithm instead of an~ad-hoc should be used.

%% Sometimes there are explosions in our case as well.
%% However it seems to be a result of a bug
%% related to the approximation of floats,
%% and e.g. NOT to proving that a more optimal plan does not exist.
%% Such bugs even depend on which compiler (or which version) is used.

\Full{
\paragraph{Profiling.}

Profiling of our tool showed
that the simulations used for collision detection and avoidance
take a~negligible part of the runtime.
Instead, most of the time is spent
in the SMT solver itself.
If our approach was applied to benchmarks
with large graphs (as discussed above),
then also encoding the formula, conversion to CNF, etc.,
would take an~additional important part of the runtime.
}

\section{\Uppercase{Conclusion}}

We have demonstrated how to solve the continuous-time MAPF problem (\MAPFR{}) by direct translation to SAT modulo linear real arithmetic. While the approach insists only on sub-optimality up to a certain factor, it shows several advantages over state-of-the-art algorithms, especially better scaling wrt. an increasing time budget and its ability to efficiently handle bottleneck situations. Our approach also allows for easy change of the objective function to another. The downside is a certain basic translation effort, especially for problems depending on large graphs.

In future work we will explore a lazy approach to translation that only generates the information necessary for solving the current problem instance. This will be especially relevant in practical applications where similar problems have to be solved repeatedly. Moreover, we will generalize the method to problems with non-linear motion functions, allowing both non-linear geometry of the involved curves and the modeling of non-linear dynamical phenomena such as acceleration of agents. The method will also benefit from the fact that the efficiency of SMT solvers is currently improving with each year.

%Future work\footnote{přidejte, co vám napadne}:
%\begin{itemize}
%    \item on-line replanning
%    \item dispatchable plans with some flexibility (temporal networks)
%    \item nonlinear curves, dynamic phenomena \cite{Kolarik:20a}
%    \item optimize for larger graphs - lazier modeling of the graph (not encoding all possibilities)
       %\item A large part of the run-time is spent on work that only depends on the given graph, but not on the start- and endpoints of the agents. Hence the time spent on the handling the graph (reading in, CNF conversion) may be re-used. However, the precise strategy needs some investigation, and we currently cannot measure the time spent in CNF conversion, hence we leave this for future work.
%    \item searching for also some plans with higher number of steps

%\end{itemize}

\ifFull
\bibliographystyle{plain}
\else
\bibliographystyle{apalike}
\fi
{\small
\bibliography{MAPF}}

\begin{thebibliography}{10}

\bibitem{AndreychukYSAS22}
Anton Andreychuk, Konstantin~S. Yakovlev, Pavel Surynek, Dor Atzmon, and Roni Stern.
\newblock Multi-agent pathfinding with continuous time.
\newblock {\em Artif. Intell.}, 305:103662, 2022.

\bibitem{AtzmonSFSK20}
Dor Atzmon, Roni Stern, Ariel Felner, Nathan~R. Sturtevant, and Sven Koenig.
\newblock Probabilistic robust multi-agent path finding.
\newblock In {\em Proc. of the Thirtieth International Conference on Automated Planning and Scheduling, 2020}, pages 29--37. {AAAI} Press, 2020.

\bibitem{Barrett:18}
Clark Barrett and Cesare Tinelli.
\newblock Satisfiability modulo theories.
\newblock In Edmund~M. Clarke, Thomas~A. Henzinger, Helmut Veith, and Roderick Bloem, editors, {\em Handbook of Model Checking}. Springer International Publishing, 2018.

\bibitem{Biere:09}
Armin Biere.
\newblock Bounded model checking.
\newblock In Biere et~al. \cite{Biere:21}.

\bibitem{Biere:21}
Armin Biere, Marijn Heule, Hans van Maaren, and Toby Walsh, editors.
\newblock {\em Handbook of Satisfiability}.
\newblock IOS Press, 2nd edition, 2021.

\bibitem{BogatarkanP019}
Aysu Bogatarkan, Volkan Patoglu, and Esra Erdem.
\newblock A declarative method for dynamic multi-agent path finding.
\newblock In {\em {GCAI} 2019. Proc. of the 5th Global Conference on Artificial Intelligence}, volume~65 of {\em EPiC Series in Computing}, pages 54--67. EasyChair, 2019.

\bibitem{Cashmore:20}
Michael Cashmore, Daniele Magazzeni, and Parisa Zehtabi.
\newblock Planning for hybrid systems via satisfiability modulo theories.
\newblock {\em Journal of Artificial Intelligence Research}, 67:235--283, 2020.

\bibitem{Cimatti:13}
Alessandro Cimatti, Alberto Griggio, Bastiaan Schaafsma, and Roberto Sebastiani.
\newblock The {MathSAT5} {SMT} solver.
\newblock In Nir Piterman and Scott Smolka, editors, {\em Proc. of TACAS}, volume 7795 of {\em LNCS}. Springer, 2013.

\bibitem{0002UKK19}
Liron Cohen, Tansel Uras, T.~K.~Satish Kumar, and Sven Koenig.
\newblock Optimal and bounded-suboptimal multi-agent motion planning.
\newblock In {\em Proc. of the Twelfth International Symposium on Combinatorial Search, {SOCS} 2019}, pages 44--51. {AAAI} Press, 2019.

\bibitem{WildeMW14}
Boris de~Wilde, Adriaan ter Mors, and Cees Witteveen.
\newblock Push and rotate: a complete multi-agent pathfinding algorithm.
\newblock {\em J. Artif. Intell. Res.}, 51:443--492, 2014.

\bibitem{FelnerSSBGSSWS17}
Ariel Felner, Roni Stern, Solomon~Eyal Shimony, Eli Boyarski, Meir Goldenberg, Guni Sharon, Nathan~R. Sturtevant, Glenn Wagner, and Pavel Surynek.
\newblock Search-based optimal solvers for the multi-agent pathfinding problem: Summary and challenges.
\newblock In {\em Proc. of the Tenth International Symposium on Combinatorial Search, {SOCS} 2017}, pages 29--37. {AAAI} Press, 2017.

\bibitem{GangeHS19}
Graeme Gange, Daniel Harabor, and Peter~J. Stuckey.
\newblock Lazy {CBS:} implicit conflict-based search using lazy clause generation.
\newblock In {\em Proc. of the Twenty-Ninth International Conference on Automated Planning and Scheduling, {ICAPS} 2019}, pages 155--162. {AAAI} Press, 2019.

\bibitem{Kolarik:23}
Tom{\'a}{\v{s}} Kol{\'a}rik and Stefan Ratschan.
\newblock Railway scheduling using boolean satisfiability modulo simulations.
\newblock In Marsha Chechik, Joost-Pieter Katoen, and Martin Leucker, editors, {\em Formal Methods}, number 14000 in LNCS, pages 56--73, Cham, 2023. Springer International Publishing.

\bibitem{LamBHS22}
Edward Lam, Pierre~Le Bodic, Daniel Harabor, and Peter~J. Stuckey.
\newblock Branch-and-cut-and-price for multi-agent path finding.
\newblock {\em Comput. Oper. Res.}, 144:105809, 2022.

\bibitem{Leofante:23}
Francesco Leofante.
\newblock Omtplan: a tool for optimal planning modulo theories.
\newblock {\em Journal on Satisfiability, Boolean Modeling and Computation}, 14(1):17--23, 2023.

\bibitem{LiCHSK21}
Jiaoyang Li, Zhe Chen, Daniel Harabor, Peter~J. Stuckey, and Sven Koenig.
\newblock Anytime multi-agent path finding via large neighborhood search.
\newblock In {\em Proc. of the Thirtieth International Joint Conference on Artificial Intelligence, {IJCAI} 2021}, pages 4127--4135. ijcai.org, 2021.

\bibitem{LunaB11}
Ryan Luna and Kostas~E. Bekris.
\newblock Push and swap: Fast cooperative path-finding with completeness guarantees.
\newblock In {\em {IJCAI} 2011, Proc. of the 22nd International Joint Conference on Artificial Intelligence, 2011}, pages 294--300. {IJCAI/AAAI}, 2011.

\bibitem{Ma22}
Hang Ma.
\newblock Intelligent planning for large-scale multi-agent systems.
\newblock {\em {AI} Mag.}, 43(4):376--382, 2022.

\bibitem{Rintanen:21}
Jussi Rintanen.
\newblock Planning and {SAT}.
\newblock In Biere et~al. \cite{Biere:21}.

\bibitem{Ryan10}
Malcolm Ryan.
\newblock Constraint-based multi-robot path planning.
\newblock In {\em {IEEE} International Conference on Robotics and Automation, {ICRA} 2010}, pages 922--928. {IEEE}, 2010.

\bibitem{Ryan08}
Malcolm R.~K. Ryan.
\newblock Exploiting subgraph structure in multi-robot path planning.
\newblock {\em J. Artif. Intell. Res.}, 31:497--542, 2008.

\bibitem{SharonSFS15}
Guni Sharon, Roni Stern, Ariel Felner, and Nathan~R. Sturtevant.
\newblock Conflict-based search for optimal multi-agent pathfinding.
\newblock {\em Artif. Intell.}, 219:40--66, 2015.

\bibitem{Silver05}
David Silver.
\newblock Cooperative pathfinding.
\newblock In {\em Proc. of the First Artificial Intelligence and Interactive Digital Entertainment Conference, 2005}, pages 117--122. {AAAI} Press, 2005.

\bibitem{Surynek19}
Pavel Surynek.
\newblock Unifying search-based and compilation-based approaches to multi-agent path finding through satisfiability modulo theories.
\newblock In {\em Proc. of the Twenty-Eighth International Joint Conference on Artificial Intelligence, {IJCAI} 2019}, pages 1177--1183. ijcai.org, 2019.

\bibitem{SurynekFSB16}
Pavel Surynek, Ariel Felner, Roni Stern, and Eli Boyarski.
\newblock Efficient {SAT} approach to multi-agent path finding under the sum of costs objective.
\newblock In {\em {ECAI} 2016 - 22nd European Conference on Artificial Intelligence}, volume 285 of {\em Frontiers in Artificial Intelligence and Applications}, pages 810--818. {IOS} Press, 2016.

\bibitem{WalkerSF18}
Thayne~T. Walker, Nathan~R. Sturtevant, and Ariel Felner.
\newblock Extended increasing cost tree search for non-unit cost domains.
\newblock In {\em Proc. of the Twenty-Seventh International Joint Conference on Artificial Intelligence, {IJCAI} 2018}, pages 534--540. ijcai.org, 2018.

\end{thebibliography}

\end{document}